\documentclass[12pt]{article}
\pdfoutput=1
\usepackage[a4paper]{geometry}
\usepackage{jheppub, amsmath,amssymb,amsfonts,amsxtra, mathrsfs, makeidx,graphics,graphicx,amsthm,epsfig, ytableau,bm, longtable,float, color,tikz,mathtools,xfrac,footnote,rotating, lscape, multirow, makecell}
\usetikzlibrary{positioning}
\usetikzlibrary{chains}
\usetikzlibrary{arrows,fit, decorations.pathreplacing}

\usepackage{bbm}

\newcommand{\ie}{\textit{i.e.}}

\numberwithin{equation}{section}

\newcommand{\be}{\begin{equation}} \newcommand{\ee}{\end{equation}}
\newcommand{\bea}{\begin{equation} \begin{aligned}} \newcommand{\eea}{\end{aligned} \end{equation}}

\def\rt2{\sqrt{2}}

\def\CH{{\cal H}}

\def\CN{{\cal N}}

\def\1{{\ds 1}}

\def\repa{\raise4pt\hbox{$\square$}\mkern-14mu\raise-4pt\hbox{$\square$}}
\def\repab{\overline{\raise4pt\hbox{$\square$}\mkern-14mu\raise-4pt\hbox{$\square$}\mkern-1mu}}

\def\smileface{\ensuremath{\hbox{\large$\bigcirc$}\mkern-15mu\raise-1pt\hbox{\scriptsize$\smallsmile$}%
\mkern-10mu\raise4pt\hbox{..}\mkern4mu}}
\def\frownface{\ensuremath{\hbox{\large$\bigcirc$}\mkern-15mu\raise-1pt\hbox{\scriptsize$\smallfrown$}%
\mkern-10mu\raise4pt\hbox{..}\mkern4mu}}

\def\node#1#2{\overset{#1}{\underset{#2}{\circ}}}
\def\sqnode#1#2{\overset{#1}{\underset{#2}{{\color{gray} \blacksquare}}}}
\def\bnode#1#2{\overset{#1}{\underset{#2}{{\color{blue} \bullet}}}}

\def\sqgrnode#1#2{\overset{#1}{\underset{#2}{{\color{gray} \blacksquare}}}}

\def\bver#1#2{\overset{{\llap{$\scriptstyle#1$}\displaystyle{\color{blue} \bullet}{\rlap{$\scriptstyle#2$}}}}{\scriptstyle\vert}}
\def\ver#1#2{\overset{{\llap{$\scriptstyle#1$}\displaystyle\circ{\rlap{$\scriptstyle#2$}}}}{\scriptstyle\vert}}

\tikzstyle{every picture}+=[remember picture]
\tikzstyle{na} = [baseline=-.5ex]

\newcommand{\ba}{\begin{array}}
\newcommand{\ea}{\end{array}}
\newcommand{\bi}{\begin{itemize}}
\newcommand{\ei}{\end{itemize}}
\def\vec#1{\bm{#1}}
\def\bea#1\eea{\allowdisplaybreaks \begin{align}#1\end{align}}
 \newcommand{\ben}{\begin{enumerate}}
\newcommand{\een}{\end{enumerate}}
\newcommand{\bean}{\begin{eqnarray*}}
\newcommand{\eean}{\end{eqnarray*}}
\newcommand{\eref}[1]{(\ref{#1})}

\newcommand{\PE}{\mathop{\rm PE}}
\newcommand{\HS}{\mathop{\rm HS}}
\newcommand{\PL}{\mathop{\rm PL}}

\newcommand{\BC}{\mathbb{C}}

\newcommand{\BZ}{\mathbb{Z}}

\newcommand{\BH}{\mathbb{H}}

\newcommand{\comment}[1]{}

\newcommand{\blue}{\color{blue}}

\newcommand{\red}{\color{red}}

\title{ 3d Coulomb branch and 5d Higgs branch at infinite coupling}
\author[a]{Giulia Ferlito}
\author[a]{, Amihay Hanany}
\author[b,c]{, Noppadol Mekareeya}
\author[d]{, and Gabi Zafrir}

\affiliation[a]{Theoretical Physics Group, Imperial College London, \\
Prince Consort Road, London, SW7 2AZ, UK}
\affiliation[b]{INFN, sezione di Milano-Bicocca, \\Piazza della Scienza 3, I-20126 Milano, Italy}
\affiliation[c]{Dipartimento di Fisica, Universit\`a di Milano-Bicocca, \\ Piazza della Scienza 3, I-20126 Milano, Italy}
\affiliation[d]{Kavli Institute for the Physics and Mathematics of the Universe, \\ University of Tokyo, Kashiwa, Chiba 277-8583, Japan}

\emailAdd{giulia.ferlito11@imperial.ac.uk}
\emailAdd{a.hanany@imperial.ac.uk}
\emailAdd{n.mekareeya@gmail.com}
\emailAdd{gabi.zafrir@ipmu.jp}

\preprint{
{\small
\begin{flushright}
IMPERIAL-TP-17-AH-08\\
\end{flushright}
}
}

\abstract{The Higgs branch of minimally supersymmetric five dimensional SQCD theories increases in a significant way at the UV fixed point when the inverse gauge coupling is tuned to zero. It has been a long standing problem to figure out how, and to find an exact description of this Higgs branch. This paper solves this problem in an elegant way by proposing that the Coulomb branches of three dimensional ${\cal N}=4$ supersymmetric quiver gauge theories, named ``Exceptional Sequences", provide the solution to the problem. Thus, once again, 3d ${\cal N}=4$ Coulomb branches prove to be useful tools in solving problems in higher dimensions. Gauge invariant operators on the 5d side consist of classical objects such as mesons, baryons and gaugino bilinears, and non perturbative objects such as instanton operators with or without baryon number. On the 3d side we have classical objects such as Casimir invariants and non perturbative objects such as monopole operators, bare or dressed. The duality map works in a very interesting way.}

\begin{document}
\maketitle

\section{Introduction and Summary}
\label{intro}
This paper is dedicated to the study of some new aspects of 5d ${\cal N}=1$ supersymmetric (with 8 supercharges) theories with $SU(n)_k$ gauge group and $N_f$ flavors, with $k$ the Chern-Simons (CS) level. At infinite coupling, for low enough $N_f$ and low enough $k$, the theory is expected to have a 5d UV fixed point with several interesting features. A well studied feature is the phenomenon of enhancement of global symmetry \cite{Seiberg:1996bd, Morrison:1996xf, Intriligator:1997pq, Aharony:1997bh, DeWolfe:1999hj, Kim:2012gu, Bashkirov:2012re, Bao:2013pwa, Bergman:2013aca, Zafrir:2014ywa, Mitev:2014jza, Tachikawa:2015mha, Hayashi:2015fsa, Yonekura:2015ksa, Gaiotto:2015una, Bergman:2015dpa} which is summarized in Table \ref{globalsym} below. A much less studied feature is the appearance of new flat directions along the Higgs branch due to the new set of massless instanton states, having a contribution to their mass given by $1/g^2$, where $g$ is the gauge coupling. These flat directions significantly increase the dimension of the Higgs branch and our goal is to get a quantitative understanding of this by developing techniques to evaluate the precise structure of ${\cal H}_\infty$, the Higgs branch at infinite coupling. The approach we take relies on pre-existing knowledge of the global symmetry at infinite coupling, $F_\infty$, and on a crucial assumption: that there exists  a 3d ${\cal N}=4$ quiver gauge theory whose Coulomb branch is precisely ${\cal H}_\infty$. Knowledge of $F_\infty$ turns out to be sufficient to fix the 3d quiver, as we demonstrate below.

Before we start describing the proposed solution, let us discuss some aspects of the problem and some known examples for the class of theories we are interested in.

The study of Higgs branches for 5d ${\cal N}=1$ $SU(n)$ theories with $N_f$ flavors already appears in \cite{Seiberg:1996bd} for the special case $n=2$. In there it is argued, via a string-theoretic analysis of the D4, D8, $\text{O8}^-$ brane system, that an $SU(2)$ gauge theory with $N_f$ flavors has a Higgs branch at infinite coupling ${\cal H}_\infty$ which is the reduced moduli space of 1 $E_{N_f+1}$ instanton on $\BC^2$  (this moduli space is nowadays \cite{Hanany:2017ooe} addressed as the closure of the minimal nilpotent orbit of $E_{N_f+1}$.

Further details on this class, in addition to the observation that for $n=2$ and $N_f=2$, namely $SU(2)$ with 2 flavors, the Higgs branch at infinite coupling is a union of two cones (see a discussion on such a feature in \cite{Ferlito:2016grh}) are provided in \cite{Morrison:1996xf} where differences in the dimensions of the Higgs branches are compared and found to be in perfect agreement with geometric data.

The special case of $SU(2)$ SYM with a trivial discrete theta angle, which displays a $E_1$ symmetry at infinite coupling, is discussed in the five brane web description of \cite{Aharony:1997ju, Aharony:1997bh}: in there, the Higgs branch at infinite coupling is realized as a separation of five branes in directions transverse to the web that open up only at infinite coupling. 

A natural generalization to the case of $SU(n)_0$ with $N_f=0$ is discussed in \cite{Cremonesi:2015lsa} where it is found that the Higgs branch at infinite coupling is ${\cal H}_\infty=\BC^2/\BZ_n$. The algebraic description of such a space is provided by operators that are relevant in this regime: the glueball operator S, an $SU(2)_R$ spin-1 object that is bilinear in the gaugino, and a set of operators known as instanton operators which are labelled $I_m$, with $m$ the instanton number, and have spin-$n |m| /2$ under $SU(2)_R$. The presence of the instanton operators at infinite coupling is expected: the previously mentioned instanton states that become massless at infinite coupling are nothing but the states created by such instanton operators. The glueball S and the instanton operators  $I_{\pm1}$ conspire to form the defining equation for $\BC^2/\BZ_n$: $S^n = I_1 I_{-1}$, with the correct $SU(2)_R$ Cartan scaling.

In view of these cases, we need a systematic analysis which allows us to determine the different contributions to ${\cal H}_\infty$. One attractive idea is to ride on the success of the study of 3d ${\cal N}=4$ Coulomb branches as spaces parametrised by a set of operators known as dressed monopole operators \cite{Cremonesi:2013lqa}. In this description, the Coulomb branch is made up of super selection sectors which are parametrized by magnetic charges which live on the GNO lattice of the gauge group. In each sector there is a special operator -- the bare monopole operator -- which carries a representation under the the R symmetry $SU(2)_R$. The representation is well defined for cases in which the Coulomb branch is a HyperK\"ahler cone. In addition there are  operators which do not change the magnetic charges, but add dressing factors to each sector with fixed magnetic charge: such operators also carry representations under $SU(2)_R$. The set of protected operators on the Coulomb branch is represented by the set of all operators from all magnetic sectors with relations that are consistent with the global symmetry and can be derived in simple cases using the algebraic properties of the moduli space. For theories with $M$ $U(1)$ factors in the gauge group, a corresponding natural topological symmetry $U(1)^M$ can be identified which takes on the role of the Cartan subalgebra of some larger, UV-hidden, symmetry group acting on the Coulomb branch. In going from the 3d UV gauge theory to the IR conformal field theory, the topological symmetry becomes thus a crucial ingredient, growing from the smallest possible Levy subgroup $U(1)^M$ to a bigger non Abelian symmetry. What about in 5d?

For each non Abelian gauge group factor, 5d theories possess a topological $U(1)$ symmetry whose associated conserved charge is known as instanton number. For the cases in this paper, there is only one gauge group factor, hence there is only one $U(1)_I$ flavor symmetry. This results in ${\cal H}_\infty$  being made of an infinite set of superselection sectors, each one parametrized by the integer instanton number. As in 3d, there is a special operator in each sector with the lowest value of spin under $SU(2)_R$. It is natural to call it the bare instanton operator (see also the discussion in \cite{Lambert:2014jna, Bergman:2016avc}). Other operators with the same instanton charge but with higher values of spin under $SU(2)_R$ can be called dressed instanton operators. In analogy to 3d, it is tempting to think of ${\cal H}_\infty$ as the space of dressed instanton operators. 

A crucial difference with the 3d case is that in 5d there is only one topological charge and the dimension of ${\cal H}_\infty$ can be very large. The dimension of the moduli space must therefore be inherited from the dimension of the space encoded in the dressing factors and the vacuum degeneracy in the bare instanton sector. This is in contrast with the 3d case where the bare sector and the classical dressing have always the same dimension. Indeed, the Coulomb branch has complex dimension equal to twice the rank of the gauge group. In 3d the number of magnetic charges is equal to the rank of the gauge group: in other words, the bare monopole operators parametrise a space of complex dimension equal to the rank. Hence the space parametrised by the dressing factors must have complex dimension equal to the rank of the gauge group. It is in fact a particularly simple space: it is freely generated by a set of Casimir invariants. On the other hand, the 5d moduli space in each instanton sector, namely the space parametrised by the classical dressing, is significantly more complicated. 

Another distinguishing feature is that monopole operators in 3d have no zero modes hence there is no degeneracy for the bare monopole operator. In contrast, the bare instanton operator has a collection of fermionic zero modes which add to its degeneracy. Such zero modes conspire to make the instanton operator transform in representations of the global symmetry which can be very large. Again a complication for the dressed instanton operator. Operators from different sectors obey some relations and these are hard to find in the absence of a physical principle. It is also crucial to point out that the 5d global symmetry evolves from being one of the largest possible Levy subgroups in the IR to the enhanced global symmetry at the UV, in contrast with the 3d case. It is therefore essential to seek for an effective description which will address all these issues. Fortunately for us, there is an answer to all these questions in the form of a 3d Coulomb branch of a different theory, which is discussed below.

The Higgs branch, being a hyperK\"ahler cone ${\cal H}$, has a natural ring structure associated with functions that are holomorphic with respect to one (of three) complex structures. This ring has a set of generators which obey transformation laws under the $SU(2)_R$ symmetry and the isometry of ${\cal H}$. For a classical Higgs branch at finite coupling, we expect the generators to be the usual objects of SQCD, the mesons at spin 1 of $SU(2)_R$ and the baryons at spin $n/2$ of $SU(2)_R$. For the special case of $n=2$ both mesons and baryons have spin 1 and the classical flavor symmetry is $SO(2N_f)\times U(1)_I$ (rather than $U(N_f)\times U(1)_I$, its maximal Levy subgroup). At infinite coupling both mesons and baryons fit into representations of the enhanced global symmetry $F_\infty$ and as a result we expect two types of instanton operators as generators of the ring. Those with spin 1 under $SU(2)_R$ which lead to symmetry enhancement and complete the mesons and gaugino bilinear into the adjoint representation of $F_\infty$, and those with spin $n/2$ which complete the baryons into a big representation of $F_\infty$. Fortunately enough, there are no additional instanton operators for the cases studied in this paper, and the generators of the ring of ${\cal H}_\infty$ transform only as spin 1 and spin $n/2$ of $SU(2)_R$. It is actually an interesting question to find a counter example to this feature where non perturbative operators have a spin under $SU(2)_R$ which is different than those which appear in the classical theory: a simple example is SYM with any gauge group \cite{Cremonesi:2015lsa}. For the special case of $n=2$ both generators again have spin 1 and following Namikawa \cite{Namikawa:2016} the moduli space is a closure of a nilpotent orbit of $F_\infty$. Indeed, the string theoretic embedding of \cite{Seiberg:1996bd} is consistent with this point, picking the minimal nilpotent orbit of $F_\infty$ as ${\cal H}_\infty$. Furthermore, for $n>2$ Namikawa's theorem implies that ${\cal H}_\infty$ is not a closure of a nilpotent orbit of $F_\infty$, but rather an extension of a nilpotent orbit by the inclusion of the generators at spin $n/2$ of $SU(2)_R$. See a recent discussion on this in \cite{Cabrera:2016vvv}.

The Coulomb branch of 3d ${\cal N}=4$ gauge theories provides a new construction of hyperK\"ahler cones. It is significantly different than the Higgs branch which is a hyperK\"ahler quotient. As a result one can obtain many new spaces using the Coulomb branch construction. A nice success is in the computation of moduli spaces of exceptional instantons as in \cite{Cremonesi:2014xha}. Another example is the construction of Higgs branches at Argyres-Douglas points \cite{Argyres:1995jj, Cecotti:2010fi} using 3d quivers \cite{Boalch2012, Xie:2012hs} as in \cite{DelZotto:2014kka}. Yet another case is the study of the Higgs branches of 6d $\CN=(1,0)$ superconformal field theories as in \cite{Mekareeya:2017jgc, Mekareeya:2017sqh}.
This paper brings another successful application of this concept as a construction of 5d Higgs branches at infinite coupling, a problem which has been standing unsolved for at least 20 years.

This brings us to the magic of hyperK\"ahler cones. They enjoy enough restrictions and enough structure to be an interesting set of moduli spaces. The global symmetry $F$ appears at spin 1 of $SU(2)_R$. This simple fact allows for exact evaluations of global symmetries in many strongly coupled theories. There is a ring structure which is ordered by the representations of $SU(2)_R$, and more importantly there is a set of generators of the ring which transform under $SU(2)_R\times F$. Namikawa's theorem \cite{Namikawa:2016} makes the set of closures of nilpotent orbits of $F$ very special, and there is a whole set of hyperK\"ahler cones which are simple extensions of these: some of them show up in this work. Perhaps these are the multiplicity free varieties, minimally unbalanced quivers, or another characterization? All these features, together with the realization of Coulomb branch global symmetries by use of Dynkin diagrams, lead to the magical results of this paper.

Below we use the features of 3d ${\cal N}=4$ theories to compute ${\cal H}_\infty$ as a Coulomb branch of a 3d quiver, ${\cal C}^{3d}$, for several families of 5d SQCD theories. Let us review the steps in finding the 3d quiver. We start from the known symmetry enhancement of the $n=2$ case. We then look for its maximal subgroups of A and D type\footnote{%
It actually remains a challenge to find cases like the $SU(9)$ maximal subgroup of $E_8$.}%
. The next step is to generalize the global symmetry to any $n$ based on a scaling of $2n$ for A type and $4n$ for D type. This gives the global symmetry enhancement for all cases studied in this paper, including the number of $U(1)$ factors in each case, thus reproducing all cases summarized in Table \ref{globalsym} using an algebraic technique rather then previous index computations. The number of $U(1)$'s tells us the number of unbalanced nodes in the quiver and we only need to figure out where in the quiver they connect. This is given by the difference of the affine Dynkin diagram of E type and the Dynkin diagram of its corresponding A or D sub algebra. With this information at hand, determining the quiver amounts to an inversion of the Cartan matrix and the solution is unique. As a non trivial test the $SU(2)_R$ spin of the operator associated to the unbalanced node, which has an imbalance of $n-2$, is given by $n/2$ and the corresponding representation under the enhanced flavor symmetry is read from the nodes attached to this unbalanced node.

Given a 3d quiver as outlined above, there are techniques which are by now standard, to evaluate the Coulomb branch of the quiver theory. One starts with a computation of the Hilbert series (HS)  \cite{Benvenuti:2006qr} using the monopole formula \cite{Cremonesi:2013lqa} (and possibly with the Hall-Littlewood formula \cite{Cremonesi:2014kwa, Cremonesi:2014vla}) in its refined version. Then one performs a character expansion and converts \cite{Hanany:2014dia} to the highest weight generating function (HWG). The results for the class of theories in this paper turn out to be particularly simple and indicates that the moduli spaces ${\cal H}_\infty$ fall into the class of multiplicity free \cite{Kobayashi:2003dv} or nearly multiplicity free varieties. This is an unexpected result and gives some hope that the moduli spaces in this paper are very simple extensions of nilpotent orbits, with a nice control over the behavior of their chiral ring. The HWG for nearly multiplicity free varieties is very simple and is given by a PE of either all positive terms (freely generated) or all positive and one negative term (complete intersection). This is a rare phenomenon which for nilpotent orbits happens for low heights only. See a set of case studies of such HWGs in \cite{Hanany:2016gbz, Hanany:2017ooe}. Is there a physical principle behind this for the class of theories of SQCD? Yet to be found. All 3d quivers in this paper are from the family of minimally unbalanced quivers. In fact, this feature is a crucial ingredient in uniquely finding all the quivers, including the three $E_4$ sequences which are particularly new. Is this a feature of SQCD? Again an interesting question which is left unanswered.

Other techniques such as finding the Plethystic Logarithm (PL) of each HS, determining the representations of the generators and the relations of the ring, getting the precise branching rules into representations of the classical flavor symmetry, and explicitly writing the classical relations and their non perturbative instanton corrections,  as carefully done in \cite{Cremonesi:2015lsa}, are left for future work.

To start the analysis we will summarize the data of the global symmetry. The following table summaries the global symmetry at infinite coupling (for $n >2$). In addition there is a summary of the main results of this paper which are given by the corresponding 3d quiver, and the HWG which encodes all the representation content of the chiral ring of ${\cal H}_\infty$ under the enhanced global symmetry.
\begin{longtable}{|c|c|c|c|c|}
\hline
\# of flavors $N_f$ & CS level $|k|$ & $F_\infty$ \thead{Global symmetry \\ at infinite coupling} & 3d quiver & HWG \\
\hline
$2n+3$ & $1/2$ & $SO(4n+8)$ & \eref{eq:2n+3} & \eref{HWGSO4np8} \\
\hline
$2n+2$ & $1$ &$SO(4n+4) \times SU(2)$ & \eref{eq:SUn1w2n+2} & \eref{HWGSO4np4SU2} \\
             & $0$ & $SU(2n+4)$ & \eref{eq:SUn0w2n+2} & \eref{HWGSU2np4} \\
\hline
$2n+1$ & $3/2$ & $SO(4n+2) \times U(1)$ & \eref{eq:SUn32w2n+1} & \eref{HWGSO4np2U1} \\
             & $1/2$ & $SU(2n+2) \times SU(2)$ & $\eref{eq:2n+1}$ & \eref{HWGSU2np2SU2} \\
\hline             
$2n$     & $2$    & $SO(4n) \times U(1)$  & \eref{eq:SUn2w2n} & \eref{HWGSO4nU1} \\
             & $1$    & $SU(2n+1) \times U(1)$ & \eref{eq:SUn1w2n} & \eref{HWGSU2np1U1} \\
             & $0$    & $SU(2n) \times SU(2) \times SU(2)$ & \eref{eq:2n}& \eref{HWGSU2np2SU2SU2} \\
\hline
$2n-1$  & $5/2$    & $SO(4n-2) \times U(1)$  & \eref{eq:SUn52w2nm1}& \eref{HWGSO4nm2U1} \\
             & $3/2$    & $SU(2n) \times U(1)$ & \eref{eq:SUn32w2nm1}& \eref{HWGSU2nU1} \\
             & $1/2$    & $SU(2n-1) \times SU(2) \times  U(1)$ & \eref{eq:2nm1}& \eref{SU2nm1SU2U1} \\
\hline
\caption{A summary of all flavor symmetry enhancements of the cases studied in this paper. The two main results of the paper are referenced by equation numbers: the 3d quiver and the highest weight generating function.}
\label{globalsym}
\end{longtable}%
Let us comment briefly about the notation of the 3d quivers. Unless specified otherwise, we denote a $U(m)$ gauge group by a circular node with a label $m$ and a bi-fundamental hypermultiplet by a line.  Let us define the imbalance of a node by the number of its flavors minus twice the number of its colors.  We refer to a node whose imbalance is non-zero as an {\it imbalanced node} and is denoted in blue.  The imbalance of all blue nodes in the 3d quivers considered in this paper is equal to $n-2$. (We return to this point below.)  Moreover, it should be emphasised that an overall $U(1)$ needs to be modded out in each of these quivers.  This can be done from any gauge node, say that with a label $p$, and the resulting gauge group of that node is $U(p)/U(1) \cong SU(p)/\BZ_p$.  In particular, when the quiver is star-shaped (\ie~ those in sections \ref{sec:2n+3} to \ref{sec:2n+1}), this is most conveniently done from the central node, as discussed in \cite{Cremonesi:2014vla}. We provide explicit examples on this in sections \ref{sec:SUn32w2n+1}, \ref{sec:SUn2w2n} and \ref{sec:2nm1}.

The flavor symmetries at infinite coupling for $N_f = 2n+3, \,\, 2n+2, \,\, 2n+1, \,\, 2n$ were presented in \cite[Table 1, p. 15]{Hayashi:2015fsa} after an intensive activity which was focused on clarifying the feature of global symmetry enhancement. For $n=2$ the flavor symmetry is enhanced to an exceptional algebra, and this can serve as a starting point for computing the corresponding 3d quiver theory. The following discussion starts with the highest possible number of flavors which admits a 5d UV fixed point, and goes down gradually with flavor number, taking into account the possible cases of CS level. It is noted that $k$ is 1/2 + integer for $N_f$ odd and an integer for $N_f$ even.

The requirement of a 5d UV fixed point sets the growth of the number of flavors to behave like $2n$. This can be viewed from the fivebrane web of the theory where there are $n$ colors given by D5 branes stretched between 2 NS branes and a set of $n$ D5 branes on each side. An increase of the number of D5 branes by two on each side by one on top and one on the bottom adds 4 more flavors. We therefore get a maximal value of $2n+4$ flavors for a theory with $k=0$. We can also determine the growth of the global symmetry in cases that it contains A type factors or D type factors. For the A type case the factor goes like $SU(2n+f)$ and for the D type case the factor goes like $SO(4n+f)$ where $f$ is an integer number which depends on $k$ and $N_f$ but not on $n$.

The case of $N_f=2n+4$ is argued to have a 6d UV fixed point \cite{Kim:2015jba, Yonekura:2015ksa} and since this paper is devoted to the study of 5d UV fixed points it is not treated here. We are led to study a one parameter family of $SU(n)$ gauge theories with $N_f=2n+3$. The level can not be 0 and we take the smallest possible value, $k=1/2$. We proceed by studying its properties and constructing ${\cal H}_\infty$, the Higgs branch at infinite coupling.

The paper is organised as follows.  In sections \ref{sec:2n+3} to \ref{sec:2nm1}, we study $SU(n)$ gauge theories with CS level $k$ and $N_f$ flavours such that $2n+3 \geq N_f \geq 2n-1$.  In each of these section, we provide the information about the global symmetry at infinite coupling, the corresponding 3d $\CN=4$ quiver and the HWG. In particular, an analysis for the case of $N_f=2n+3$ is discussed in great detail in section \ref{sec:2n+3}.  In section \ref{sec:5dweb}, we realize the connection between the 5d theories of our interest and the corresponding 3d quiver using the web of fivebranes. 

\section{$SU(n)_{\pm1/2}$ with $N_f=2n+3$ flavors, $E_8$ sequence} \label{sec:2n+3}
For $n=2$, the highest number of flavors with a 5d UV fixed point is 7. The global symmetry at infinite coupling is $F_\infty = E_8$. It is therefore suggestive to call this family the $E_8$ sequence. As above, the CS level is $k= \frac{1}{2}$. Higher values of $k$ for generic values of $n$ are argued not to have a 5d UV fixed point. The Chern Simons density vanishes for $n=2$, hence the value of $k$ is insignificant for this case. The Higgs branch at infinite coupling is identified as the reduced moduli space of the 1 $E_8$ instanton on $\BC^2$, and there is a natural 3d quiver with such a Coulomb branch - the affine $E_8$ Dynkin diagram. We write the following relation

\subsection{$n=2$}
The Higgs branch at infinite coupling is given by the Coulomb branch of the $3d$ quiver:\\
\bea 
\CH_\infty \left(\node{}{SU(2)}-\sqgrnode{}{SO(14)} \right)= \mathcal{C}^{3d} \left ( \node{}{1}-\node{}{2}-\node{}{3}-\node{}{4}-\node{}{5}-\node{\ver{}{3}}{6}-\node{}{4}-\node{}{2}  \tikz[na]; \right ) \eea
i.e. ${\cal H}_\infty$ is the closure of the minimal nilpotent orbit of $E_8$, where $\CH_\infty$ denotes the Higgs branch of the 5d theory at infinite coupling, and $\mathcal{C}^{3d}$ denotes the Coulomb branch of the 3d ${\cal N}=4$ theory. Higher values of $n$ in Table \ref{globalsym} reveal that the global symmetry is $SO(4n+8)$, corresponding to $SO(16)$ for $n=2$. Indeed $SO(16)$ is a maximal subgroup of $E_8$ and an inspection of the quiver reveals that there is a special node of label 2 at the right of the quiver that singles out when one takes the difference between the affine $E_8$ Dynkin diagram and the $SO(16)$ Dynkin diagram. This node is connected to the $SO(16)$ spinor node reflecting the fact that the adjoint representation of $E_8$ is decomposed into the adjoint representation and one of the spinor representations of $SO(16)$. Our goal is to construct a one parameter family of quivers with a global symmetry $SO(4n+8)$ such that for $n=2$ it reduces to the affine $E_8$ quiver. To do so, we recall a crucial conjecture on the global symmetry of 3d ${\cal N}=4$ Coulomb branches:

\paragraph{Global symmetry conjecture.} Introduce the notion of a balanced node by setting the number of its flavors to equal twice its rank;
then the subset of balanced nodes forms the Dynkin diagram of the non Abelian factor of the global symmetry on the Coulomb branch. The number of $U(1)$ factors is the number of unbalanced nodes minus 1. This conjecture can be extended to cases of non simply laced Dynkin diagrams by replacing the balance condition by using the Cartan matrix. Set the flavors of the quiver to form a vector $f$ and the ranks to form a vector $r$, Set the Cartan matrix to be $A$ and the balance condition is given by $A r = f$. Given two roots $\alpha, \beta$ with monopole operators $V_{m_1}$ and $V_{m_2}$, respectively, with spin 1 under $SU(2)_R$, it is straightforward to show that if $\alpha+\beta$ is a root, then $V_{m_1+m_2}$ also has spin 1 under $SU(2)_R$. There are exceptions to this conjecture with low rank non simply laced factors in the global symmetry, where this criterion gives rise to a simply laced maximal sub algebra of the flavor symmetry, but these cases do not show up in the study of this paper.

We note that for the present case there are no $U(1)$ factors in the global symmetry, leading to 1 unbalanced node, and this node is connected to the spinor node, in order to match to the $n=2$ case. This fixes the family uniquely and is presented below.
\subsection{$E_8$ Sequence, $n>2$}
The theory is $SU(n)$ with $N_f=2n+3$ flavors and $k=\frac{1}{2}$.
The Higgs branch at infinite coupling is given by the Coulomb branch of the $3d$ quiver:\\
\bea \label{eq:2n+3}
\CH_\infty \left(\node{}{SU(n)_{\pm\frac{1}{2}}} - \sqgrnode{}{U(2n+3)} \right)= \mathcal{C}^{3d} \left ( \node{}{1}-\cdots -\node{}{2n+1}-\node{\ver{}{n+1}}{2n+2}-\node{}{n+2}-\bnode{}{2}  \tikz[na]; \right )~. \eea
From which the dimension can be derived simply,
\be
\dim \CH_\infty = 2n^2+7n+7
\ee
 This increases the finite coupling Higgs branch which has dimension $(2n+3)n-(n^2-1)=n^2+3 n+1$ by adding $n^2+4n+6$ new flat directions.
 All the nodes are balanced except for the last one on the right. Henceforth unbalanced nodes are depicted in blue. There is a classical global symmetry $U(2n+3) \times U(1)_I \cong SU(2n+3) \times U(1)_B \times U(1)_I $ with rank $r=2n+4$. The global symmetry conjecture implies that the 
flavor symmetry at infinite coupling is $SO(4n+8)$, which has the same rank as the finite coupling global symmetry. 
 
Indeed, for $n=2$, by our construction, we verify that there is a larger set of balanced nodes, resulting in a global symmetry $E_8 \supset SO(16)$.
 
Let us now concentrate on the last node on the right hand side. it has an imbalance of $N_f-2N_c=n-2$. Consequently, the lowest $SU(2)_R$ spin for a 3d monopole operator with non-zero fluxes associated to this last gauge node is $n/2$. Such a monopole transforms in the spinor representation of $SO(4n+8)$.
Thus the chiral ring at infinite coupling is generated by an $SO(4n+8)$ adjoint rep at $SU(2)_R$ spin-1 and an $SO(4n+8)$ spinor rep at $SU(2)_R$ spin-$(n/2)$. 

The resulting highest weight generating function (HWG) is 
\be \label{HWGSO4np8}
\PE \left[ \sum_{i=1}^{n+1} \mu_{2 i} t^{2 i}+ t^4 + \mu_{2 n + 4} (t^n + t^{n + 2}) \right]~.
\ee
The special case of $n=2$ was studied in \cite[Table 10, p.41]{Hanany:2015hxa} (the row containing $E_8$ and $D_8$). The case of general $n$ was discussed in \cite[Eq. (26)]{Hanany:2017pdx} with their $N$ being our $n+1$.  A simple observation of this HWG reveals that the lattice of weights for this moduli space consists of the adjoint and one of the spinor representations, but not the other 2 sub lattices of $SO(4n+8)$. This situation resembles the case of the perturbative spectrum of the Heterotic $SO(32)$ string where the gauge group is sometimes said to be $Spin(32)/\BZ_2$.

From \eref{HWGSO4np8}, we compute the Hilbert series and plethystic logarithms for the cases of $n=2,\, 3, \,4$ to the power of $t^{2n}$ as follows:
\be
\begin{split}
\HS_{n=2} &= 1+ ([0,1,0,0,0,0,0,0]+[0,0,0,0,0,0,0,1])t^2 + (1 + [0, 0, 0, 0, 0, 0, 0, 1]  \\
& \qquad + [0, 0, 0, 0, 0, 0, 0, 2] + [0, 0, 0, 1, 0, 0, 0, 0] + [0, 1, 0, 0, 0, 0, 0, 1] \\
& \qquad + [0, 2, 0, 0, 0, 0, 0, 0]) t^4+ \ldots \\
\HS_{n=3} &= 1+[0,1,0,0,0,0,0,0,0,0] t^2 + [0, 0, 0, 0, 0, 0, 0, 0, 0, 1]t^3 \\
& \qquad + ([0, 0, 0, 0, 0, 0, 0, 0, 0, 0] + [0, 0, 0, 1, 0, 0, 0, 0, 0, 0]  \\
& \qquad + [0, 2, 0, 0, 0, 0, 0, 0, 0, 0])t^4 + ([0, 0, 0, 0, 0, 0, 0, 0, 0, 1]  \\
& \qquad + [0, 1, 0, 0, 0, 0, 0, 0, 0, 1])t^5+ ([0, 0, 0, 0, 0, 0, 0, 0, 0, 2] + [0, 0, 0, 0, 0, 1, 0, 0, 0, 0]  \\
& \qquad  + [0, 1, 0, 0, 0, 0, 0, 0, 0, 0] + [0, 1, 0, 1, 0, 0, 0, 0, 0, 0] + [0, 3, 0, 0, 0, 0, 0, 0, 0, 0])t^6 \\
& \qquad +\ldots \\
\HS_{n=4} &= 1+ [0, 1, 0, 0, 0, 0, 0, 0, 0, 0, 0, 0] t^2 +(1+ [0, 0, 0, 0, 0, 0, 0, 0, 0, 0, 0, 1]    \\
& \qquad + [0, 0, 0, 1, 0, 0, 0, 0, 0, 0, 0, 0]+ [0, 2, 0, 0, 0, 0, 0, 0, 0, 0, 0, 0])t^4 \\
& \qquad  + ([0, 0, 0, 0, 0, 0, 0, 0, 0, 0, 0, 1] + [0, 0, 0, 0, 0, 1, 0, 0, 0, 0, 0, 0] \\
& \qquad  + [0, 1, 0, 0, 0, 0, 0, 0, 0, 0, 0, 0] + [0, 1, 0, 0, 0, 0, 0, 0, 0, 0, 0, 1] \\
& \qquad  + [0, 1, 0, 1, 0, 0, 0, 0, 0, 0, 0, 0] + [0, 3, 0, 0, 0, 0, 0, 0, 0, 0, 0, 0])t^6 \\
& \qquad  + (1 + [0, 0, 0, 0, 0, 0, 0, 0, 0, 0, 0, 1] + [0, 0, 0, 0, 0, 0, 0, 0, 0, 0, 0, 2] \\
& \qquad + [0, 0, 0, 0, 0, 0, 0, 1, 0, 0, 0, 0] + [0, 0, 0, 1, 0, 0, 0, 0, 0, 0, 0, 0] \\
& \qquad + [0, 0, 0, 1, 0, 0, 0, 0, 0, 0, 0, 1]+  [0, 0, 0, 2, 0, 0, 0, 0, 0, 0, 0, 0] \\
& \qquad  + [0, 1, 0, 0, 0, 0, 0, 0, 0, 0, 0, 1]+ [0, 1, 0, 0, 0, 1, 0, 0, 0, 0, 0, 0]  \\
& \qquad + [0, 2, 0, 0, 0, 0, 0, 0, 0, 0, 0, 0]+ [0, 2, 0, 0, 0, 0, 0, 0, 0, 0, 0, 1]  \\
& \qquad + [0, 2, 0, 1, 0, 0, 0, 0, 0, 0, 0, 0] + [0, 4, 0, 0, 0, 0, 0, 0, 0, 0, 0, 0]) t^8 + \ldots~,
\end{split}
\ee
and
\be
\begin{split}
\PL_{n=2} &= ([0,1,0,0,0,0,0,0]+[0,0,0,0,0,0,0,1])t^2 - ([2,0,0,0,0,0,0,0]  +[1, 0, 0, 0, 0, 0, 1, 0] \\
& \qquad+[0,0,0,1,0,0,0,0]+1) t^4 + \ldots~,
\end{split}
\ee
\be
\begin{split}
\PL_{n=3} &= [0,1,0,0,0,0,0,0,0,0] t^2 + [0, 0, 0, 0, 0, 0, 0, 0, 0, 1]t^3 -[2, 0, 0, 0, 0, 0, 0, 0, 0, 0] t^4\\ 
& \qquad - [1,0,0,0,0,0,0,0,1,0]t^5 + ( [2,0,0,0,0,0,0,0,0,0]- [0,0,0,0,0,1,0,0,0,0] \\
& \qquad -[0,1,0,0,0,0,0,0,0,0])t^6  + \ldots~,
\end{split}
\ee
\be
\begin{split}
\PL_{n=4} &= [0, 1, 0, 0, 0, 0, 0, 0, 0, 0, 0, 0] t^2 + ([0,0,0,0,0,0,0,0,0,0,0,1] \\ 
&\qquad - [2,0,0,0,0,0,0,0,0,0,0,0]) t^4+ ([2,0,0,0,0,0,0,0,0,0,0,0] \\
&\qquad -[1,0,0,0,0,0,0,0,0,0,1,0])t^6 +([1, 0, 0, 0, 0, 0, 0, 0, 0, 0, 1, 0] \\
&\qquad + [2, 0, 0, 0, 0, 0, 0, 0, 0, 0, 0, 1]  - [0, 0, 0, 0, 0, 0, 0, 1, 0, 0, 0, 0] \\ 
&\qquad - [0, 0, 0, 1, 0, 0, 0, 0, 0, 0, 0, 0]  - [0, 1, 0, 0, 0, 0, 0, 0, 0, 0, 0, 0] -1) t^8 + \ldots~.
\end{split}
\ee

This computation leads to the general behavior of the generators and relations for any $n$.
\begin{enumerate}
\item At order 2 there is one generator in the representation $[0,1,0,\ldots,0]$. Let $A^{ij}$ for $i,j=1\ldots4n+8$ be the chiral ring operators.
\item At order $n$ there is one generator in the representation $[0,\ldots,0,1]$. Let $\Psi_\alpha$ for $\alpha=1\ldots 2^{2n+3}$ be the chiral ring operators.
\item At order 4 there is one relation in the representation $[2,0,\ldots,0]$. It is simple to write down, $A^{ij} A^{jk} = \delta^{ik} Tr(A^2)$.
\item At order $n+2$ there is one relation in the representation $[1,0,\ldots,0,1,0]$. $A^{ij} \gamma^{j}_{\alpha \dot \alpha}\Psi_\alpha = 0$, where the $\alpha$ contraction is symmetric for $n$ even and antisymmetric for $n$ odd. This relation removes points from the other spinor sub lattice.
\item At order $2n$ there are several antisymmetric relations ($\frac{n+2}{2}$ for $n$ even and $\frac{n+1}{2}$ for $n$ odd) in the representation $\wedge^k$, for $k<2n+4$ which satisfies $k=2$ mod 4 for $n$ odd and $k=0$ mod 4 for $n$ even. $\left(A^n\right)^{[i_1 \cdots i_{k}]} = \Psi_\alpha \gamma^{i_1\ldots i_{k}}_{\alpha \beta}\Psi_\beta$.
\end{enumerate}

\subsubsection{Special cases}
It is instructive to look at special cases as they teach us several interesting physical properties of these moduli spaces. For $n=1$ the 5d theory is trivial and we expect the 3d quiver to reflect this. It takes the form
\bea 
\mathcal{C}^{3d} \left ( \node{}{1}-\node{}{2}-\node{}{3}-\node{\ver{}{2}}{4}-\node{}{3}-\bnode{}{2}  \tikz[na]; \right ) = \BH^{16} \eea
with a trivial moduli space of dimension 16. This is verified by the generators of $SU(2)_R$ spin $1/2$ transforming in the spinor representation of $SO(12)$ of dimension 32. The Higgs branch is generated by 32 free half hypers $\Psi_\alpha$ and the remaining relations indicate that the moduli space is indeed freely generated.
At order 2 we get that the adjoint representation is no longer a generator of the moduli space
\be
A^{ij} = \Psi_\alpha \gamma^{ij}_{\alpha\beta}\Psi_\beta
\ee
where the spinor indices are contracted with an epsilon.
At order 3 we get a relation which vanishes by the properties of the Clliford algebra.
\be
A^{ij} \gamma^{j}_{\alpha \dot \alpha}\Psi_\alpha = \Psi_\alpha \gamma^{ij}_{\alpha\beta}\Psi_\beta \gamma^{j}_{\gamma \dot \alpha}\Psi_\gamma = 0.
\ee

For $n=2$ the spinor of $SO(16)$ has spin 1 under $SU(2)_R$ and we get the expected enhancement of the global symmetry to $E_8$. All relations are now at order 4, combining into the Joseph relations for the $E_8$ algebra, as expected from the closure of the minimal nilpotent orbit of $E_8$. Explicitly, the Joseph relations transform in the $[1,0,0,0,0,0,0,0]+[0,0,0,0,0,0,0,0]$ representation of $E_8$ of dimensions $3875 +1$ respectively. These representations decompose to $[2,0,0,0,0,0,0,0]+[1,0,0,0,0,0,1,0]+[0,0,0,1,0,0,0,0]+[0,0,0,0,0,0,0,0]$ of $SO(16)$ of dimensions $135+ 1920+1820+1$, respectively. The singlet relation sets the second Casimir invariant of $E_8$ to 0, which in turn sets the second Casimir invariant of $SO(16)$ to be proportional to the quadratic invariant of the spinor of $SO(16)$.


\subsubsection{$SO(4n+8) ~\longrightarrow ~SU(2n+3) \times U(1) \times U(1)$}

Next we decompose such representations of $SO(4n+8)$ into those of $SU(2n+3) \times U(1) \times U(1)$.  Since there are two $U(1)$'s involved, we fix the linear combinations of them, called $U(1)_I$ and $U(1)_B$, by the following conditions.
\bi
\item At $SU(2)_R$ spin-1, we decompose the adjoint representation of $SO(4n+8)$ to those of $SU(2n+3) \times U(1)_B \times U(1)_I$.  We require that the operators have integer instanton numbers. This results in a charge assignment that sets the rank-2 antisymmetric representation $\wedge^2 = [0,1,\ldots,0]$  of $SU(2n+3)$ to carry $U(1)_I$ charge $1$, and those in the fundamental representation $\wedge^1 = [1,0,\ldots,0]$ of $SU(2n+3)$ to carry $U(1)_I$ charge $2$.
\item At $SU(2)_R$ spin-$(n/2)$, we decompose the spinor representation $[0,\ldots,0,1]$ of $SO(4n+8)$ to those of $SU(2n+3) \times U(1)_B \times U(1)_I$.   We require that the operators in the $\wedge^n = [0,\ldots,0,1,0,\ldots,0]$ representation of $SU(2n+3)$ carry $U(1)_B$ charge $n$, and those in $\wedge^{n+1}$ of $SU(2n+3)$ carry $U(1)_B$ charge $-\frac{1}{2}$.
\ei
These two conditions fix $U(1)_B$ and $U(1)_I$ charges for all $SU(2n+3)$ representations resulting from the decompositions.   The generating functions for such decompositions are
\be
\text{[0,\ldots,0,1]:} \begin{cases} 
b^{\frac{1}{4} (2 n+3) (n-1)} q^{-\frac{1}{2} (n+3)} \frac{1+b  q^2  \wedge}{1- q b^{\frac{1}{2}-n} \wedge ^2} &\qquad  \text{$n$ odd}  \\
b^{\frac{1}{4} (2 n+3) n} q^{-\frac{n}{2}} \frac{1+b^{-n-\frac{1}{2}} q^{-1} \wedge}{1-q b^{\frac{1}{2}-n} \wedge ^2 } &\qquad  \text{$n$ even} 
\end{cases}
\ee
with $\wedge^{2n+4} = 0$.
The following decomposition of the vector representation of $SO(4n+8)$ proves to be useful
\be
[1,0, \ldots, 0] \quad \rightarrow \quad \wedge^1_{\frac{1-2n}{4},\frac{1}{2}} + \wedge^{2n+2}_{\frac{2n-1}{4},-\frac{1}{2}}  + \wedge^0_{\frac{2n+3}{4}, \frac{3}{2}} + \wedge^0_{-\frac{2n+3}{4}, -\frac{3}{2}} 
\ee
It gives
\be
\label{2.14}
[0,1,0, \ldots, 0] \quad \rightarrow \quad \left( \wedge^2_{\frac{1-2n}{2}, 1} + \wedge^1_{1,2} + \wedge^1_{-\frac{2n+1}{2},-1} + \text{c.c.} \right) + {\bf adj}_{0,0} +\wedge^0_{0,0} + \wedge^0_{0,0}
\ee
\be
[0,0,0,1,0, \ldots, 0] \quad \rightarrow \quad \left( \wedge^4_{{1-2n}, 2} + \wedge^1_{1,2} + \wedge^1_{-\frac{2n+1}{2},-1} + \text{c.c.} \right) + {\bf adj}_{0,0} +\wedge^0_{0,0} + \wedge^0_{0,0}
\ee

As an example, we summarize the result for $n=3$ below:
\be \label{decompSU2np3}
\begin{split}
[0,1,0, \ldots, 0] \quad &\rightarrow \quad \left( \wedge^1_{-\frac{7}{2},-1} + \wedge^1_{1,2}  + \wedge^2_{-\frac{5}{2}, 1}+ \text{c.c.} \right) + {\bf adj}_{0,0} +\wedge^0_{0,0} + \wedge^0_{0,0} ~, \\
[0, \ldots, 0,1] \quad &\rightarrow \quad \wedge^0_{\frac{9}{2}, -3}  + \wedge^1_{\frac{11}{2}, -1}  + \wedge^2_{2, -2}+  \wedge^3_{3,0} + \wedge^4_{-\frac{1}{2}, -1}  + \text{c.c.} ~,
\end{split}
\ee
where $\wedge^k$ denotes the rank-$k$ antisymmetric representation of $SU(9)$; the subscript denote the charges under $U(1)_B \times U(1)_I$; and `c.c.' denotes the conjugate representations with the opposite $U(1)_{B,I}$ charges of what have been written before.

The right hand sides of \eref{decompSU2np3} consist of mesons, the gaugino bilinear, instantons, baryons and baryonic instantons, whose
 transformations properties are tabulated in \eref{tab:transfGenereators}.
\be \label{tab:transfGenereators}
\begin{tabular}{|c@{\hskip 0.2cm}|c@{\hskip 0.2cm}|c@{\hskip 0.2cm}|c@{\hskip 0.2cm}|c@{\hskip 0.2cm}|}
\hline
& mesons &  baryons & \begin{tabular}{@{}c@{}} baryonic instantons  \end{tabular}  & gaugino bilinear  \\
\hline
$SU(2)_R $ &  spin-1 &  $n/2$ & $n/2$ & 1  \\
\hline
 \begin{tabular}{@{}c@{}} $SU(2n+3) \times$ \\ $ U(1)_B \times U(1)_I$ \end{tabular}  & $ {\bf adj}_{0,0} + \wedge^0_{0,0}$ &  $\wedge^n_{n,0}+\wedge^{n+3}_{-n,0}$ & $\wedge^p_{b,i}$ ~\text{with $b, i \neq 0$}  & $\wedge^0_{0,0}$  \\
\hline
\end{tabular}
\ee

\subsection{5d analysis}

It is interesting to compare the results we get from the 3d quiver against direct computations in the 5d gauge theory. We will not perform an extensive 5d analysis but rather study some of the simpler BPS objects of the theory, particularly those constructed from perturbative states and 1-instanton contributions. The latter can be computed using the methods of \cite{Tachikawa:2015mha,Zafrir:2014ywa,Yonekura:2015ksa}. Let's first consider the structure of the SCFT Higgs branch. While here we considered the SCFT as the UV completion of the 5d $SU(n)_{\pm \frac{1}{2}}+(2n+3)F$ gauge theory, it has additional dual descriptions, that is other 5d gauge theories that have that SCFT as their UV completion. One that is quite useful for our purposes is the 5d $USp(2n-2)+(2n+3)F$ gauge theory \cite{GK}. Here the perturbative global symmetry is $SO(4n+6) \times U(1)_I$ and it is conjectured to enhance to $SO(4n+8)$ by 2-instanton particle contributions \cite{BZ}. The theory also receives traceable contributions from the 1-instanton. These are in the $\bold{n+1}$ of $SU(2)_R$ and in the spinor of $SO(4n+6)$. The instanton and anti-instanton contributions then form one state in the $\bold{n+1}$ of $SU(2)_R$ and in a chiral spinor of the enhanced $SO(4n+8)$. This precisely matches the spinor we observe from the 3d quiver.

Next we return to the 5d gauge theory $SU(n)_{\pm \frac{1}{2}}+(2n+3)F$. We can perform a similar analysis also on it. Particularly the 1-instanton sector receives three contributions, two of which leads to additional conserved currents. The exact charges depend on the sign of the Chern-Simons level, where here we take the minus sign. In that case one finds the 1-instanton contributes conserved currents in the antisymmetric and anti-fundamental of $SU(2n+3)$ with baryon charges $-n+\frac{1}{2}$ and $n+\frac{1}{2}$ respectively. The anti-instanton provides the complex conjugate. This motivates the decomposition we used for $SO(4n+8)$ in equation \ref{2.14}.

There is one more 1-instanton contribution which is in the $\bold{n+1}$ of $SU(2)_R$ and in the $(n+2)^\mathrm{th}$ antisymmetric representation of $SU(2n+3)$, with the anti-instanton contributing the complex conjugate. These are readily identified as part of the spinor representation. The perturbative matter appearing in \eref{tab:transfGenereators}, can also be readily identified with the analogous states in the gauge theory. So we see that we can at least map the states identifiable in the 5d gauge theory to those we observe from the 3d quiver\footnote{When performing the spinor decomposition one finds that sometimes there are additional 1-instanton contributions besides the one specified here. This happens already at $n=4$. These states appear to be gauge valued 1-instanton states whose gauge charges are canceled by perturbative matter. These won't appear in the method that we used as it only observes gauge invariant 1-instanton states. These can in principle be seen from more sophisticated methods like the 5d superconformal index. At least for $n=4$, we have indeed checked that these can be reproduced using this method.}. 

\section{$SU(n)_{\pm1}$ with $N_f=2n+2$ flavors, $E_7$ sequences} \label{sec:SUn1w2n+2}
To proceed with a lower number of flavors, we recall that for $n=2$ the global symmetry is $E_7$. Its algebra has two maximal subgroups with A or D type factors. Correspondingly, there are two distinct $E_7$ sequences. The first in our discussion is $SO(12)\times SU(2) \subset E_7$. The case of $SU(8) \subset E_7$ is dealt in the next section. Fitting $SO(12)\times SU(2)$ with a scaling of $4n$ and 12 for $n=2$ gives a global symmetry of $SO(4n+4)\times SU(2)$, as expected from the global symmetry for this series. There is a special node which is found by computing the difference between the affine $E_7$ Dynkin diagram and the $SO(12)\times SU(2)$ Dynkin diagram. This is the node which connects one of the spinor nodes of the D Dynkin diagram to the $SU(2)$ node. As there are no $U(1)$ factors in the global symmetry we expect one unbalanced node and identify it with the special node. This uniquely fixes the 3d quiver as below.

The Higgs branch at infinite coupling is given by the Coulomb branch of the $3d$ quiver:\\
\bea \label{eq:SUn1w2n+2}
\CH_\infty \left(\node{}{SU(n)_{\pm1}}-\sqgrnode{}{U(2n+2)} \right)= \mathcal{C}^{3d} \left ( \node{}{1}-\cdots -\node{}{2n-1}-\node{\ver{}{n}}{2n}-\node{}{n+1}-\bnode{}{2}-\node{}{1}  \tikz[na]; \right )~. 
\eea
\be \label{dimSO4np4SU2}
\dim \CH_\infty = 2n^2+3n+3
\ee
This increases the finite coupling Higgs branch that has dimension $(2n+2)n-(n^2-1)= n^2+2n+1$ by adding $n^2+n+2$ new flat directions.

All the nodes are balanced except for node $\bnode{}{2}$. There is a classical global symmetry $U(2n+2) \times U(1)_I \cong SU(2n+2) \times U(1)_B \times U(1)_I $ with rank $r=2n+3$. 
The global symmetry is computed from the quiver, by removing the unbalanced node: it gives a flavor symmetry $SO(4n+4) \times SU(2)$ at infinite coupling, which has rank $2n+3$, the same as the finite coupling global symmetry.

The HWG is
\be \label{HWGSO4np4SU2}
\PE \left[ \sum _{i=1}^n \mu _{2 i} t^{2 i} +t^4+\nu ^2 t^2+\nu  \mu _{2 n+2} \left(t^{n}+t^{n+2}\right) 
+\mu _{2 n+2}^2 t^{2 n+2} 
-\nu ^2 \mu _{2 n+2}^2 t^{2 n+4} \right]
\ee
where $\mu_k$ keeps track of the highest weight of $SO(4n+4)$ and $\nu$ keeps track of the highest weight of $SU(2)$. The special case of $n=2$ was studied in \cite[Table 10, p.41]{Hanany:2015hxa}; the row containing $E_7$ and $D_6 \otimes A_1$. 

Another non-trivial test of \eref{HWGSO4np4SU2} is to derive the dimension of $\CH_\infty$ from this function, in the way described in \cite[sec. 4.3]{Hanany:2015hxa}, and compare it with \eref{dimSO4np4SU2}. To derive the former, we use the following data: 
\bi
\item The HWG dimension of \eref{HWGSO4np4SU2} is $n+5-1 = n+4$.  
\item The irrep structure of $SO(4n+4) \times SU(2)$ that appears in the HWG is $[0,m,0,m,\ldots,0,m]_{SO(4n+4)}[m]_{SU(2)}$, with $n\neq 0$.  The dimension of such a representation is a polynomial in $m$ of degree $4 n^2+5n+2$.
\ei
The sum of the above two quantities is $4n^2+6n+6$.  This is the expected complex dimension of $\CH_\infty$ as derived from the conjectured HWG. Indeed, it is in agreement with the quaternionic dimension given by \eref{dimSO4np4SU2}.  The conjectured HWG thus passes this test.

\subsection{5d analysis}

We can again compare some of the states observed from the 3d quiver with direct analysis in 5d. Specifically for the SCFT Higgs branch we can again use a dual description of the SCFT, $USp(2n-2)+(2n+2)F$ gauge theory, to study some of the states on the Higgs branch. In this theory the classical global symmetry is $SO(4n+4) \times U(1)_I$ and it is argued to enhance to $SO(4n+4) \times SU(2)$ by 2-instanton particle contributions \cite{BZ}. We again have contributions from the 1-instanton in the $\bold{n+1}$ of $SU(2)_R$ and in the spinor of $SO(4n+4)$. Now the instanton and anti-instanton states form a doublet of the enhanced $SU(2)$. This leads to a state on the Higgs branch in the $\bold{n+1}$ of $SU(2)_R$ and in the $(\bold{2^{2n+1}},\bold{2})$ of $SO(4n+4) \times SU(2)$. This matches the state we see coming from the unbalanced node in the 3d quiver.

\section{$SU(n)_0$ with $N_f=2n+2$ flavors} \label{sec:2n+2}
This corresponds to the second $E_7$ sequence.
\subsection{$n=2$}
The gauge theory is $SU(2)$ with $N_f=6$ flavors and it flows from a SCFT at infinite coupling which displays symmetry enhancement. The Higgs branch is indeed given by the Coulomb branch of the 3d quiver:\\
\bea 
\CH_\infty \left(\node{}{SU(2)}-\sqgrnode{}{SO(12)} \right)= \mathcal{C}^{3d} \left ( \node{}{1}-\node{}{2}-\node{}{3}-\node{\ver{}{2}}{4}-\node{}{3}-\node{}{2}-\node{}{1}  \tikz[na]; \right ) \eea
which has global symmetry $E_7$. Such a global symmetry can be evinced by recalling that the above is precisely the \emph{affine} Dynkin diagram for $E_7$. The Higgs branch dimension  at infinite coupling is 
\be
\dim \CH_\infty= 17
\ee
This increases the finite coupling Higgs branch that has dimension $12-3=9$ by adding $8$ new flat directions.

\subsection{$n>2$}
The 5d theory is $SU(n)_0$ with $N_f=2n+2$ flavors. We proceed as before, by looking at the maximal sub algebra $SU(8)\subset E_7$ which generalizes from $n=2$ with a scaling of $2n$ to $SU(2n+4)$. This global symmetry coincides with the expected global symmetry for this sequence. The 3d quiver has only one unbalanced node since there are no $U(1)$ factors in the global symmetry. Furthermore the node which is unbalanced is attached to the middle node of the $SU(2n+4)$ Dynkin diagram, as this is the extra node for the $SU(8)$ Dynkin diagram inside $E_7$. These points fix the 3d quiver uniquely as below.
\bea  \label{eq:SUn0w2n+2}
\CH_\infty \left(\node{}{SU(n)_0}-\sqgrnode{}{U(2n+2)} \right)= \mathcal{C}^{3d} \left ( \node{}{1}-\cdots -\node{}{n+1}-\node{\bver{}{2}}{n+2}-\node{}{n+1}-\cdots-\node{}{1}  \tikz[na]; \right )~. 
\eea
\be
\dim \CH_\infty = n^2+4n+5
\ee
This increases the finite coupling Higgs branch that has dimension $n^2+2n+1$ by adding $2n+4$ new flat directions.

There is a classical global symmetry $U(2n+2) \times U(1)_I \cong SU(2n+2) \times U(1)_B \times U(1)_I $ with rank $r=2n+3$. 
All the nodes are balanced except for the node at the top. The global symmetry can be read from the quiver, after removing the unbalanced node: it gives a flavour symmetry $SU(2n+4)$ at infinite coupling, which has rank $2n+3$, the same as the finite coupling global symmetry. 

For $n=2$ there is again a further enhancement, since the infinite coupling global symmetry is $E_7 \supset SU(8)$.
 
The top node has again an imbalance of $N_f-2N_c=n-2$. Analogously to the previous case, there is a 3d monopole operator at $SU(2)_R$ spin-$n/2$. Here the unbalanced node is connected to the $(n+2)^\mathrm{th}$ node, hence the monopole transforms in the $(n+2)^\mathrm{th}$ antisymmetric representation of $SU(2n+4)$.

Thus the chiral ring at infinite coupling is generated by two $SU(2n+4)$ reps: the adjoint at $SU(2)_R$ spin-1 and the $(n+2)^\mathrm{th}$  antisymmetric rep at $SU(2)_R$ spin-$(n/2)$. 

The HWG is given by
\be\label{HWGSU2np4}
\PE \left[ \sum _{i=1}^{n+1} \mu_i  \mu_{2 n+4-i} t^{2 i}+t^4+\mu_{n+2} \left(t^n+t^{n+2}\right) \right]
\ee
The special case of $n=2$ was studied in \cite[Table 10, p.41]{Hanany:2015hxa}; the row containing $E_7$ and $A_7$. The cases of $n=3$ and $n=4$ were also studied in \cite[secs. 5.4, 5.6]{Hanany:2017pdx}.

\subsection{5d analysis}

Once again we compare the structure inferred from the 3d quiver against analysis in 5d. Now we shall employ the direct theory that is $SU(n)_0 + (2n+2)F$. Performing 1-instanton analysis on this theory one finds three contributions. Two give additional conserved currents and, together with expected 2-instanton contributions, should enhance the classical $SU(2n+2)\times U(1)_B \times U(1)_I$ to the $SU(2n+4)$ observed in the 3d quiver. 

The third one is in the $\bold{n+1}$ of $SU(2)_R$, has zero baryonic charge, and in the $(n+1)^\mathrm{th}$ antisymmetric representation of $SU(2n+2)$. We have one from the instanton and one from the anti-instanton. Additionally we have the perturbative baryons and ant-baryons, also in the $\bold{n+1}$ of $SU(2)_R$, but in the $n^\mathrm{th}$ and $(n+2)^\mathrm{th}$ antisymmetric representations of $SU(2n+2)$ respectively. These four contributions merge to form the $(n+2)^\mathrm{th}$ antisymmetric representation of the enhanced $SU(2n+4)$, again in agreement with what is observed in the 3d quiver.

\section{$SU(n)_{\pm3/2}$ with $N_f=2n+1$ flavors, $E_6$ sequences} \label{sec:SUn32w2n+1}
For this case, the $n=2$ global symmetry is $E_6$. We proceed as before by finding its maximal subgroups with A or D type factors, We find $SO(10)\times U(1)$ and $SU(6)\times SU(2)$. The second case is dealt in the next section, The first has an expected scaling of $4n$ and sets the global symmetry to be $SO(4n+2)\times U(1)$. Hence the 3d quiver has 2 unbalanced nodes. By looking at the difference between the affine $E_6$ Dynkin diagram and the $SO(10)$ Dynkin diagram we determine these 2 nodes. These points determine the quiver uniquely, as below.
The Higgs branch at infinite coupling is given by the Coulomb branch of the $3d$ quiver:\\
\bea \label{eq:SUn32w2n+1}
\CH_\infty \left( \node{}{SU(n)_{\pm\frac{3}{2}}}-\sqnode{}{U(2n+1)}  \right) = \mathcal{C}^{3d} \left ( \node{}{1}-\cdots-\node{}{2n-2} -\node{\overset{\bver{}{1}}{\ver{}{n}}}{2n-1}-\node{}{n}-\bnode{}{1} \tikz[na]; \right )~. \eea
In fact the star-shaped quiver diagram on the right hand side was studied in \cite[sec. 4.2.1]{Hanany:2017pdx}.  The quaternionic dimension of the Higgs branch at infinite coupling is
\be
\dim \CH_\infty = 2n^2+n+1
\ee
This increases the finite coupling Higgs branch that has dimension $(2n+1)n-(n^2-1)=n^2+n+1$ by adding $n^2$ new flat directions.

By setting $n=2$ we confirm that the global symmetry reduces to $E_6$.
The global symmetry at infinite coupling can be read off from the quiver, after removing the blue nodes: this operation results with a flavor symmetry $SO(4n+2) \times U(1)$, which has rank $2n+2$, preserving the rank of the symmetry at finite coupling.

As pointed out in \cite[Eq. (25)]{Hanany:2017pdx} (with their $N$ being $n-1$), the HWG is given by
\be \label{HWGSO4np2U1}
\begin{split}
	\PE \left[ \sum _{i=1}^{n-1} \mu _{2 i} t^{2 i} +t^2 + \left(\mu _{2 n} q +\frac{\mu _{2 n+1}}{q}\right)t^n
	\right]~.
\end{split}
\ee	
The $n=2$ case was studied in \cite[(2.110)]{Cremonesi:2015lsa}.  

For $n=3$, it can be shown that the Hilbert series obtained from \eref{HWGSO4np2U1} agrees with the Coulomb branch Hilbert series of the quiver in the right hand side of \eref{eq:SUn32w2n+1}.  The Coulomb branch Hilbert series for general $n$ can be obtained using the Hall-Littlewood formula \cite{Cremonesi:2014kwa, Cremonesi:2014vla}, which involves ``gluing'' the Hilbert series of $T_{[1^{2n-1}]}(SU(2n-1))$, $T_{[(n-1)^2,1]}(SU(2n-1))$ and $T_{[(n-1)^2,1]}(SU(2n-1))$ via the common symmetry $U(2n-1)/U(1)$. Indeed, in this case, an overall $U(1)$ symmetry in the 3d quiver can be conveniently modded out from the central node.  Explicitly the Hilbert series for $n=3$ is given by (see \cite[(3.29)]{Cremonesi:2014vla})
\be
\begin{split} 
&\mathrm{HS}_{n=3}(t; \vec x, \vec y, \vec z) \\
&= \sum_{n_1 \geq n_2 \geq n_3  \geq n_4 \geq {\blue n_5 =0}}  \left\{ t^{-2\delta_{U(5)}(n_1, \ldots, n_5)} P_{U(5)}(t; n_1, \ldots, n_5) \right\} {\blue (1-t^2)} \\
& \qquad \times H[T_{[1^{5}]}(SU(5))](t; x_1,\ldots, x_5; n_1, \ldots, n_5) H[T_{[2^2,1]}(SU(5))](t; y_1,\ldots, y_5;  n_1, \ldots, n_5)  \\
& \qquad \times H[T_{[2^2,1]}(SU(5))](t; z_1,\ldots, z_5;  n_1, \ldots, n_5)~. 
\end{split} 
\ee
where $H[T_{\vec \rho}(SU(N))]$ is the Coulomb branch Hilbert series of $T_{\vec \rho}(SU(N))$ given by (3.9) of \cite{Cremonesi:2014kwa} and the factor $t^{-2\delta_{U(m)}(n_1, \ldots, n_m)} P_{U(m)}(t; n_1, \ldots, n_m)$ denotes the contribution of the vector multiplet of $U(m)$ group with $P_{U(m)}(t; n_1, \ldots, n_m)$ given by (A.2) of \cite{Cremonesi:2013lqa} and $\delta_{U(m)}(n_1, \ldots, n_m) = \sum_{1\leq i < j \leq m} (n_i -n_j)$.\footnote{It should be noted that the parameter $t$ in \cite{Cremonesi:2013lqa, Cremonesi:2014kwa} should be replaced by $t^2$ to conform with the convention in this paper.}  The factor in the curly brackets denotes a gauging of the $U(5)$ symmetry and the terms in blue denote the removal of an overall $U(1)$ from the $U(5)$ node\footnote{See the discussion around (3.3) of \cite{Cremonesi:2014vla}.}; hence this amounts to taking the central node of the 3d quiver in \eref{eq:SUn32w2n+1} (labelled by $2n-1$) to be $U(5)/U(1) \cong SU(5)/\BZ_5$.  Setting all $x_i$, $y_i$ and $z_i$ to 1, we obtain
\be
1+92 t^2+128 t^3+4173 t^4+9984 t^5+127920 t^6+ \ldots~,
\ee
in agreement with \eref{HWGSO4np2U1}.
\subsection{5d analysis}

This 5d SCFT also has a $USp(2n-2)$ gauge theory dual, now with $2n+1$ flavors. In fact from this description the classical and quantum symmetries match. The 1-instanton spectrum is, as before, made of one state in the $\bold{n+1}$ of $SU(2)_R$ and in the spinor of $SO(4n+2)$. These exactly reproduce the contributions from the two unbalanced nodes, one from the instanton and the other from the anti-instanton, where we remind the reader that spinors of $SO(4n+2)$ are complex.

\section{$SU(n)_{\pm1/2}$ with $N_f=2n+1$ flavors} \label{sec:2n+1}
This is the second $E_6$ sequence.
\subsection{$n=2$}
The gauge theory is $SU(2)$ with $N_f=5$ flavors. The Higgs branch at infinite coupling is given by the Coulomb branch of the 3d quiver:\\
\bea 
\CH_\infty \left( \node{}{SU(2)}-\sqnode{}{SO(10)}  \right) = \mathcal{C}^{3d} \left ( \node{}{1}-\node{}{2}-\node{\overset{\ver{}{1}}{\ver{}{2}}}{3}-\node{}{2}-\node{}{1}  \tikz[na]; \right ) \eea
which is the affine Dynkin diagram for $E_6$. This is precisely the global symmetry of the Higgs branch at infinite coupling.  The Higgs branch dimension  at infinite coupling is 
\be
\dim \CH_\infty= 11
\ee
This increases the finite coupling Higgs branch that has dimension $10-3=7$ by adding $4$ new flat directions.

\subsection{$n>2$}
The 5d theory is $SU(n)_{\pm1/2}$ with $N_f=2n+1$ flavors. For this case we look at the maximal subgroup $SU(6)\times SU(2)$ of $E_6$. The scaling determines the global symmetry to be $SU(2n+2)\times SU(2)$, indicating that there is only one unbalanced node and the difference between the affine $E_6$ Dynkin diagram and the $SU(6)\times SU(2)$ Dynkin diagram sets this node to connect the middle rank antisymmetric node to the $SU(2)$ node. These points determine the quiver uniquely as below.  The Higgs branch at infinite coupling is given by the Coulomb branch of the $3d$ quiver:\\
\bea  \label{eq:2n+1}
\CH_\infty \left( \node{}{SU(n)_{\frac{1}{2}}}-\sqnode{}{U(2n+1)}  \right) = \mathcal{C}^{3d} \left ( \node{}{1}-\cdots-\node{}{n}-\node{\overset{\ver{}{1}}{\bver{}{2}}}{n+1}-\node{}{n}-\cdots-\node{}{1} \tikz[na]; \right )~. \eea
\be
\dim \CH_\infty = n^2+2n+3
\ee
This increases the finite coupling Higgs branch that has dimension $n^2+n+1$ by adding $n+2$ new flat directions.

There is a classical global symmetry $U(2n+1) \times U(1)_I \cong SU(2n+1) \times U(1)_B \times U(1)_I $ with rank $r=2n+2$. 

The global symmetry at infinite coupling can be read off from the quiver, after cutting the vertical leg: this operation results with a flavor symmetry $SU(2) \times SU(2n+2)$, which has rank $2n+2$, preserving the rank of the symmetry at finite coupling. 

The imbalance of the lowest of the two vertical nodes is given by $N_f-2N_c=n-2$. The 3d monopole operators carrying the smallest flux under this node carry spin-$(n/2)$ and transform in the (spin-$1/2$, $(n+1)$-antisymmetric) of $SU(2) \times SU(2n+2)$.

As it is by now clear, the $n=2$ case is special. Spin-$n/2$ is in this case spin-1, i.e there are extra operators at spin-1 and thus the enhancement of the symmetry is larger. Indeed $E_6 \supset SU(2) \times SU(6)$.

The HWG is given by
\be \label{HWGSU2np2SU2}
\begin{split}
 \PE \left[ \sum _{i=1}^{n+1} \mu _i \mu _{2 n+2-i} t^{2 i} +\nu ^2 t^2 +t^4  +\nu  \mu _{n+1} \left(t^n+t^{n+2}\right) -\nu ^2 \mu^2 _{n+1} t^{2 n+4} \right]~.
\end{split}
\ee
The cases of $n=2$ and $n=3$ were studied in \cite[Tab. 10, p. 41, row 1]{Hanany:2015hxa} and \cite[Eq.(55)]{Hanany:2017pdx}, respectively.

\subsection{5d analysis}

In this case we do not have a $USp$ dual so we analyze the direct case. There are again three contributions at the 1-instanton order, two in the $\bold{3}$ of $SU(2)_R$, corresponding to conserved currents, and one in the $\bold{n+1}$ of $SU(2)_R$. The former are in the fundamental and a singlet of $SU(2n+1)$ and together with the anti-instanton contribution enhance the classical $SU(2n+1) \times U(1)_B \times U(1)_I$ to $SU(2) \times SU(2n+2)$.

 The last state is in the $(n+1)^\mathrm{th}$ antisymmetric representation of $SU(2n+1)$ (here we have chosen the minus sign for the CS level), and we also have the anti-instanton in the $n^\mathrm{th}$ antisymmetric representation. We also have the baryons and anti-baryons contributing with the same $SU(2)_R$ and $SU(2n+1)$ representations. These merge with the corresponding instanton to form a doublet of the enhanced $SU(2)$, while the $n^\mathrm{th}$ and $(n+1)^\mathrm{th}$ antisymmetric representations of $SU(2n+1)$ form the $(n+1)^\mathrm{th}$ antisymmetric representation of the enhanced $SU(2n+2)$. Overall we see that the 1-instantons, baryons and their conjugates exactly form the state expected from the unbalanced node in the 3d quiver.

\section{$SU(n)_{\pm2}$ with $N_f=2n$ flavors, $E_5$ sequences} \label{sec:SUn2w2n}
For $n=2$ the global symmetry is $E_5 = SO(10)$. It has 3 maximal subgroups with A or D type factors given by $SO(8)\times U(1)$, $SU(5) \times U(1)$, and $SU(4)\times SU(2)\times SU(2)$. Correspondingly, there are 3 different $E_5$ sequences. For the first case, the global symmetry generalizes to $SO(4n)\times U(1)$, implying two unbalanced nodes in the 3d quiver. The difference between the affine $E_5$ Dynkin diagram and the $SO(8)$ Dynkin diagram identifies these nodes, and the quiver is again fixed uniquely by these points. 
The Higgs branch at infinite coupling is given by the Coulomb branch of the $3d$ quiver:\\
\bea \label{eq:SUn2w2n}
\CH_\infty \left( \node{}{SU(n)_{2}}-\sqnode{}{2n}  \right) 
=\mathcal{C}^{3d} \left ( \node{}{1}-\node{}{2}-\cdots-\node{}{2n-3}-\node{\ver{n-1}{}}{2n-2}-\node{\bver{}{1}}{n}-\bnode{}{1} \right )~. 
\eea
\be
\dim \CH_\infty = 2n^2-n+1
\ee
This increases the finite coupling Higgs branch that has dimension $2n^2-(n^2-1)=n^2+1$ by adding $n^2-n$ new flat directions.

The global symmetry at infinite coupling can be read off from the above quiver, after removing the blue nodes: this operation results with a flavor symmetry $SO(4n) \times U(1)$, which has rank $2n+1$, as expected from finite coupling. The HWG takes the form
\be \label{HWGSO4nU1}
\begin{split}
\PE \left[ \sum _{i=1}^{n-1} \mu _{2 i} t^{2 i} +t^2 +\left(q+\frac{1}{q}\right) \mu _{2 n} t^n \right]~.
\end{split}
\ee
The case of $n=2$ was studied in \cite[(2.88)]{Cremonesi:2015lsa}.  For $n=3$, it can be shown that the Hilbert series obtained from \eref{HWGSO4nU1} agrees with the Coulomb branch Hilbert series of the quiver in the right hand side of \eref{eq:SUn2w2n}.  The latter can be obtained using the Coulomb branch formula \cite{Cremonesi:2013lqa} and the Hall-Littlewood formula \cite{Cremonesi:2014kwa, Cremonesi:2014vla}. Explicitly the Hilbert series for $n=3$ is given by
\be
\begin{split} 
	&\mathrm{HS}_{n=3}(t; \vec x, \vec y, \vec z, \vec w) \\
	&= \sum_{m_1 \geq \cdots \geq m_4 > -\infty }\,\, \sum_{n_1 \geq n_2 \geq {\blue n_3 =0}}\,\, H[T_{[1^{4}]}(SU(4))](t; x_1,\ldots, x_4; m_1, \ldots, m_4) \\
	& \qquad \times H[T_{[2^2]}(SU(4))](t; y_1,\ldots, y_4; m_1, \ldots, m_4)  \times {\red t^{\sum_{i=1}^4\sum_{j=1}^3 |m_i -n_j|}}  \\
	& \qquad \times H[T_{[2,1]}(SU(3))](t; z_1,z_2, z_3; n_1,n_2,n_3)  H[T_{[2,1]}(SU(3))](t; w_1,w_2, w_3; n_1,n_2,n_3) \\
	& \qquad \times \left\{ t^{-2\delta_{U(4)}(m_1, \ldots, m_4)} P_{U(4)}(t; m_1, \ldots, m_4) \right\} \\
	& \qquad \times \left\{ t^{-2\delta_{U(3)}(n_1, \ldots, n_3)} P_{U(3)}(t; n_1, \ldots, n_3) \right\}  {\blue (1-t^2)}~. 
\end{split} 
\ee
where $H[T_{\vec \rho}(SU(N))]$ is the Coulomb branch Hilbert series of $T_{\vec \rho}(SU(N))$ given by (3.9) of \cite{Cremonesi:2014kwa} and the factor $t^{-2\delta_{U(m)}(n_1, \ldots, n_m)} P_{U(m)}(t; n_1, \ldots, n_m)$ denotes the contribution of the vector multiplet of the $U(m)$ group with $P_{U(m)}(t; n_1, \ldots, n_m)$ given by (A.2) of \cite{Cremonesi:2013lqa} and $\delta_{U(m)}(n_1, \ldots, n_m) = \sum_{1\leq i < j \leq m} (n_i -n_j)$.\footnote{It should be noted that the parameter $t$ in \cite{Cremonesi:2013lqa, Cremonesi:2014kwa} should be replaced by $t^2$ to conform with the convention in this paper.}  The factor in red denotes the contribution from the hypermultiplet in the bi-fundamental representation of $U(4) \times U(3)$. The first and the second pairs of the curly brackets denote gauging of the $U(4)$ and $U(3)$ symmetries respectively. The terms in blue denote the removal of an overall $U(1)$ from the $U(3)$ node\footnote{See the discussion around (3.3) of \cite{Cremonesi:2014vla}.}. Hence this amounts to taking the node labelled by $n$, which connects to the two blue nodes, to be $U(3)/U(1) \cong SU(3)/\BZ_3$.  Setting all $x_i$, $y_i$, $z_i$ and $w_i$ to 1, we obtain
\be
1 + 67 t^2 + 64 t^3 + 2200 t^4 + 3520 t^5 + 47707 t^6+\ldots~.
\ee
Alternatively, we can remove an overall $U(1)$ from the node labelled by $2n-2$ by changing the summations to be
\be
\sum_{m_1 \geq \cdots \geq m_3 \geq {\blue m_4 =0} }\,\,\,\, \sum_{n_1 \geq n_2 \geq  n_3 > -\infty}~.
\ee
This yields the same Hilbert series as above.

\subsection{5d analysis}

This 5d SCFT also has a $USp(2n-2)$ gauge theory dual, now with $2n$ flavors, where again in this description the classical and quantum symmetries match. The 1-insanton spectrum is, as before, made of one state in the $\bold{n+1}$ of $SU(2)_R$ and in the spinor of $SO(4n)$. These exactly reproduce the contributions from the two unbalanced nodes, one from the instanton and the other from the anti-instanton, where we remind the reader that spinors of $SO(4n)$ are self-conjugate.

\section{$SU(n)_{\pm1}$ with $N_f=2n$ flavors} \label{sec:SUn1w2n}
This is the second $E_5$ sequence.

Here the corresponding subgroup of $E_5$ is $SU(5)\times U(1)$ which generalizes to $SU(2n+1)\times U(1)$. This indicates that there are two unbalanced nodes in the 3d quiver which are determined by looking at the difference between the affine $E_5$ Dynkin diagram and the $SU(5)$ Dynkin diagram. These points determine the 3d quiver uniquely.
The Higgs branch at infinite coupling is given by the Coulomb branch of the $3d$ quiver:\\
\bea \label{eq:SUn1w2n}
\CH_\infty \left( \node{}{SU(n)_{1}}-\sqnode{}{2n}  \right) 
=\mathcal{C}^{3d} \left ( \node{}{1}-\node{}{2}-\cdots-\node{}{n-1}-\node{\bver{1}{}}{n}-\node{\bver{}{1}}{n}-\node{}{n-1}- \cdots -\node{}{1} \right)~. 
\eea
\be
\dim \CH_\infty = n^2+n+1
\ee
This increases the finite coupling Higgs branch that has dimension $n^2+1$ by adding $n$ new flat directions.

The global symmetry at infinite coupling can be read off from the above quiver, after removing the blue nodes: this operation results with a flavor symmetry $SU(2n+1) \times U(1)$, which has rank $2n+1$, as expected from finite coupling.

The HWG is
\be \label{HWGSU2np1U1}
\begin{split}
\PE \left[\sum_{i=1}^{n-1} \mu_i \mu_{2n-i+1} t^{2i} + t^2 + (\mu_{n} q + \mu_{n+1}q^{-1}) t^n\right]
\end{split}
\ee
For $n=2$, we recover the Hilbert series of the minimal nilpotent orbit of $E_5=SO(10)$ written in terms of the highest weight of $SU(5)\times U(1)$.  For $n=3$, it can be shown that the Hilbert series obtained from \eref{HWGSU2np1U1} agrees with the Coulomb branch Hilbert series of the quiver in the right hand side of \eref{eq:SUn1w2n}.  The latter can be obtained using the Coulomb branch formula \cite{Cremonesi:2013lqa} and the Hall-Littlewood formula \cite{Cremonesi:2014kwa, Cremonesi:2014vla}.  The unrefined Hilbert series for $n=3$ is
\be
1+49 t^2+70 t^3+1176 t^4+2716 t^5+19452 t^6+\ldots~.
\ee

\subsection{5d analysis}

In this case we do not have a $USp$ dual so we analyze the direct case. There are now two contributions at the 1-instanton order, one of which is a conserved current in the fundamental of $SU(2n)$, that together with the anti-instanton leads to the enhancement of the symmetry to $SU(2n+1)$. The second one is in the $\bold{n+1}$ of $SU(2)_R$ and the $(n+1)^\mathrm{th}$ antisymmetric representation of $SU(2n)$. It, together with the anti-instanton, baryons and anti-baryons form the states expected from the unbalanced nodes in the 3d quiver.

\section{$SU(n)_0$ with $N_f=2n$ flavors} \label{sec:2n}
This is the third $E_5$ sequence. 

\subsection{$n=2$}
The gauge theory is $SU(2)$ with $N_f=4$ flavors. The Higgs branch at infinite coupling is given by the Coulomb branch of the 3d quiver:\\
\bea 
\CH_\infty \left( \node{}{SU(2)}-\sqnode{}{SO(8)}  \right) = \mathcal{C}^{3d} \left ( \node{}{1}- \node{\overset{{\llap{$\overset{{}}{\overset{{1}}{\circ}} -$}\displaystyle \overset{{2}}\circ{\rlap{$-\overset{{}}{\overset{{1}}{\circ}}$}}}}{\scriptstyle\vert}}{2} -\node{}{1} \tikz[na];  \right )~. \eea \\
This is the affine Dynkin diagram for $SO(10)$. The Coulomb branch of this quiver is known to correspond to the moduli space of one $SO(10)$ instanton, namely the minimal nilpotent orbit of $SO(10)$.  

The Higgs branch dimension at infinite coupling is 
\be
\dim \CH_\infty= 7
\ee
This increases the finite coupling Higgs branch that has dimension $8-3=5$ by adding $2$ new flat directions.

\subsection{$n>2$}
The theory is $SU(n)_0$ with $N_f=2n$ flavors. The corresponding subgroup of $E_5$ is $SU(4)\times SU(2)\times SU(2)$, which generalizes to $SU(2n)\times SU(2)\times SU(2)$. This implies that there is only 1 unbalanced node and it is given by the difference between the affine $E_5$ Dynkin diagram and the $SU(4)\times SU(2)\times SU(2)$ Dynkin diagram. These points determine the 3d quiver uniquely. 
The Higgs branch at infinite coupling is given by the Coulomb branch of the $3d$ quiver:\\
\bea \label{eq:2n}
\CH_\infty \left( \node{}{SU(n)_{0}}-\sqnode{}{2n}  \right) =\mathcal{C}^{3d} \left ( \node{}{1}-\node{}{2}-\cdots- \node{}{n-1}-\node{\overset{{\llap{$\overset{{}}{\overset{{1}}{\circ}} -$}\displaystyle \overset{{2}}{{\blue \bullet}}{\rlap{$-\overset{{}}{\overset{{1}}{\circ}}$}}}}{\scriptstyle\vert}}{n} -\node{}{n-1}-\cdots -\node{}{2}-\node{}{1} \tikz[na];  \right )~. \eea
\be
\dim \CH_\infty = n^2+3
\ee
This increases the finite coupling Higgs branch that has dimension $n^2+1$ by adding $2$ new flat directions.

There is a classical global symmetry $U(2n) \times U(1)_I \cong SU(2n) \times U(1)_B \times U(1)_I $ with rank $r=2n+1$. 
The global symmetry at infinite coupling can be read off from the above quiver, after cutting off the subquiver connected to the node with label $n$: this operation results on a flavor symmetry $SU(2) \times SU(2) \times  SU(2n)$, which has rank $2n+1$, as expected from finite coupling.

The unbalanced node is the middle node on the top line of the quiver and again has an imbalance of $N_f-2N_c=n-2$. Similarly to the previous cases, the monopole operator with non-zero flux under this gauge group has  $SU(2)_R$ spin-$n/2$ and transforms in the ($1/2$, $1/2$, $n$-antisymmetric) of $SU(2) \times SU(2) \times SU(2n+2)$.

As expected, for $n=2$, there is a further symmetry enhancement: indeed  $SO(10) \supset SU(2) \times SU(2) \times SU(4)$.

The highest weight generating function is given by
\be \label{HWGSU2np2SU2SU2}
\begin{split}
	\PE \left[ \sum _{i=1}^n \mu _i \mu _{2 n-i} t^{2 i} +\left(\nu _1^2+\nu _2^2\right) t^2 +t^4 +\nu _1 \nu _2 \mu _n \left(t^n+t^{n+2}\right) -\nu _1^2 \nu _2^2 \mu _n^2 t^{2 n+4} \right]~.
\end{split}
\ee
The special case of $n=2$ was considered in \cite[Tab. 9, p. 40]{Hanany:2015hxa} in the row containing $D_5$ and $D_3 \otimes D_2$, with $m_1=\mu_2$, $m_2 = \mu_1$, $m_3 = \mu_3$, and $n_i =\nu_i$.  For $n=3$, we check up to order $t^6$ that the Hilbert series obtained from \eref{HWGSU2np2SU2SU2} is in agreement with the Coulomb branch Hilbert series of the 3d quiver on the right hand side of \eref{eq:2n}.  The latter can be computed using a mixture of the Coulomb branch formula \cite{Cremonesi:2013lqa} and the Hall-Littlewood formula \cite{Cremonesi:2014kwa, Cremonesi:2014vla}. The unrefined Hilbert series is
\be
1 + 41 t^2 + 80 t^3 + 824 t^4 + 2560 t^5 + 12434 t^6 + \ldots .
\ee

\subsection{5d analysis}

We can again study the 1-instanton spectrum of this theory in 5d. There are three contributions at the 1-instanton order. The first two are singlets of $SU(2n)$, carry baryonic charge of $\pm \frac{n}{2}$ and are conserved currents \cite{Tachikawa:2015mha}. Together with the anti-instanton, these lead to the enhancement of the symmetry to $SU(2)^2$, and this also explains the decomposition we have chosen. The second one is in the $\bold{n+1}$ of $SU(2)_R$, carries zero baryonic charge and is in the $n^\mathrm{th}$ antisymmetric representation of $SU(2n)$. Once again, when combined with the anti-instanton, baryons and anti-baryons, these exactly form the states we observed from the unbalanced node in the 3d quiver.

\section{$SU(n)_{\pm5/2}$ with $N_f=2n-1$ flavors, $E_4$ sequences} \label{sec:SUn52w2nm1}
For $n=2$ the global symmetry is $E_4=SU(5)$. There are 3 subgroups of $SU(5)$ with A or D factors. They are $SO(6)\times U(1)$, $SU(4)\times U(1)$, and $SU(3)\times SU(2)\times U(1)$\footnote{The first two are the same as Lie algebras, but they are naturally associated with different generalizations.}. Correspondingly, there are 3 different $E_4$ sequences. We discuss the first. The $SO(6)\times U(1)$ global symmetry generalizes to $SO(4n-2)\times U(1)$, implying that there are two unbalanced nodes in the 3d quiver. These are identified by looking at the difference between the affine $E_4$ Dynkin diagram and the $SO(6)$ Dynkin diagram. This information fixes the Dynkin diagram uniquely which is given below.
The Higgs branch at infinite coupling is given by the Coulomb branch of the $3d$ quiver:\\
\bea \label{eq:SUn52w2nm1}
\CH_\infty \left( \node{}{SU(n)_{\pm\frac{5}{2}}}-\sqnode{}{2n-1}  \right)  = \mathcal{C}^{3d} \left ( \begin{tikzpicture}[baseline={([yshift=-.5ex]current bounding box.center)}]
\node[anchor=south west] at (0,0)  {$\node{}{1}- \node{}{2}-\cdots- \node{}{2n-4}-\node{\overset{\bver{}{1}}{\ver{}{n-1}}}{2n-3}-\node{}{n-1}-\bnode{}{1}$};   \draw[black] (4.5,1.6)--(6.2,0.7); 
\end{tikzpicture} \right )~.  \eea
\be \label{dimSO4nm2U1}
\dim \CH_\infty = 2n^2-3n+2
\ee
This increases the finite coupling Higgs branch that has dimension $(2n-1)n-(n^2-1)=n^2-n+1$ by adding $n^2-2n+1$ new flat directions.

There is a classical global symmetry $U(2n-1) \times U(1)_I \cong SU(2n-1) \times U(1)_B \times U(1)_I $ with rank $r=2n$.
The global symmetry at infinite coupling can be read off from the above quiver, after removing the blue nodes: this operation results with a flavor symmetry $SO(4n-2) \times  U(1)$, which has rank $2n$, as expected from finite coupling.

The HWG is 
\be \label{HWGSO4nm2U1}
\begin{split}
 \PE\left[ \sum _{i=1}^{n-2} \mu _{2 i} t^{2 i}+t^2+t^n \left(q \mu _{2 n-2}+\frac{\mu _{2 n-1}}{q}\right)+\mu _{2 n-2} \mu _{2 n-1} \left(t^{2 n-2}-t^{2 n}\right)  \right] .
\end{split}
\ee
Indeed, for $n=2$, we recover the Hilbert series of the minimal nilpotent orbit of $E_4 = SU(5)$, written in terms of the highest weight of $SO(6)\times U(1)$.

Another non-trivial test of \eref{HWGSO4nm2U1} is to derive the dimension of $\CH_\infty$ from this function, in the way described in \cite[sec. 4.3]{Hanany:2015hxa}, and compare it with \eref{dimSO4nm2U1}. To derive the former, we use the following data: 
\bi
\item The HWG dimension of \eref{HWGSO4nm2U1} is $(n-2)+4-1 = n+1$.  
\item The irrep structure of $SO(4n-2)$ that appears in the HWG is $[0,m,0,m,\ldots,0,m,0,m,m]$, with $m\neq 0$.  The dimension of such a representation is a polynomial in $m$ of degree $4 n^2-7 n+3$.
\ei
The sum of the above two quantities is $4n^2-6n+4$.  This is the expected complex dimension of $\CH_\infty$ as derived from the conjectured HWG. Indeed, it is in agreement with the quaternionic dimension given by \eref{dimSO4nm2U1}.  The conjectured HWG thus passes this test.

\subsection{5d analysis}

Again the analysis in 5d is easiest in the dual $USp(2n-2)+(2n-1)F$ frame. The 1-instanton contribution are just as before, the instanton and anti-instanton, both in the $\bold{n+1}$ of $SU(2)_R$ and in the spinor and its conjugate of $SO(4n-2)$. These indeed match the contributions of the two unbalanced nodes in the 3d quiver.

\section{$SU(n)_{\pm3/2}$ with $N_f=2n-1$ flavors} \label{sec:SUn32w2nm1}
This is the second $E_4$ sequence.

For this case, the corresponding subgroup of $E_4$ is $SU(4)\times U(1)$ which generalizes to $SU(2n)\times U(1)$. This implies that there are 2 unbalanced nodes which can be determined by looking at the difference between the affine $E_4$ Dynkin diagram and the $SU(4)$ Dynkin diagram. This fixes the 3d quiver uniquely.
\bea \label{eq:SUn32w2nm1}
\CH_\infty \left( \node{}{SU(n)_{\pm\frac{3}{2}}}-\sqnode{}{2n-1}  \right)  = \mathcal{C}^{3d} \left ( \begin{tikzpicture}[baseline={([yshift=-.5ex]current bounding box.center)}]
\node[anchor=south west] at (0,0)  {$\node{}{1}- \node{}{2}-\cdots- \node{}{n-2}-\node{\bver{1}{}}{n-1}-\node{}{n-1}-\node{\bver{}{1}}{n-1}- \node{}{n-2} -\node{}{2}-\node{}{1}$} ;   \draw[black] (4.1,1.15)--(5.95,1.15); 
\end{tikzpicture} \right )~.  \eea
The quaternionic Higgs branch dimension at infinite coupling of the quiver on the left hand side can be computed from the Coulomb branch dimension of the quiver on the right hand side:
\be \label{dimSU2nU1}
\dim \CH_\infty = n^2
\ee
This increases the finite coupling Higgs branch that has dimension $n^2-n+1$ by adding $n-1$ new flat directions.

The global symmetry at infinite coupling can be read off from the above quiver, after removing the blue nodes: this operation results with a flavor symmetry $SU(2n) \times  U(1)$, which has rank $2n$, as expected from finite coupling.

The HWG is conjectured to be
\be \label{HWGSU2nU1}
\PE \left[ \sum_{i=1}^{n-1} \mu_i \mu_{2n-i} t^{2i} +t^2 + (\mu_{n-1}q + \mu_{n+1} q^{-1})t^n - \mu_{n-1} \mu_{n+1}t^{2n} \right]~.
\ee
Indeed, for $n=2$, we recover the Hilbert series of the minimal nilpotent orbit of $E_4 = SU(5)$, written in terms of the highest weight of $SU(4)\times U(1)$. For $n=3$, we compute the Coulomb branch Hilbert series of the quiver of the right hand side of \eref{eq:SUn32w2nm1} up to order $t^6$ using the Coulomb branch formula \cite{Cremonesi:2013lqa}, and find the agreement with the Hilbert series obtained from the HWG \eref{HWGSU2nU1}.  In which case, the unrefined Hilbert series is
\be
1 + 36 t^2 + 30 t^3 + 630 t^4 + 798 t^5 + 7210 t^6+ \ldots~,
\ee
where more details of the computation will be given in the next section.

Another non-trivial test of \eref{HWGSU2nU1} is to derive the dimension of $\CH_\infty$ from this function, in the way described in \cite[sec. 4.3]{Hanany:2015hxa}, and compare it with \eref{dimSU2nU1}. To derive the former, we use the following data: 
\bi
\item The HWG dimension of \eref{SU2nm1SU2U1} is $(n-1)+1+2-1 = n+1$.  
\item The irrep structure of $SU(2n)$ that appears in the HWG is $[m,\cdots,m,0,m, \cdots, m]$, with $m\neq 0$.  The dimension of such a representation is a polynomial in $m$ of degree ${2n \choose 2}-1 = 2n^2-n-1$.
\ei
The sum of the above two quantities is $2n^2$.  This is the expected complex dimension of $\CH_\infty$ as derived from the conjectured HWG. Indeed, it is in agreement with the quaternionic dimension given by \eref{dimSU2nU1}.  The conjectured HWG thus passes this test.

\subsection{5d analysis}

We can again study the 1-instanton spectrum of this theory in 5d. There are two contributions at the 1-instanton order. The first is in the fundamental of $SU(2n-1)$ and, together with the anti-instanton, provides the conserved currents that enhances $SU(2n-1)$ to $SU(2n)$. The second one is in the $\bold{n+1}$ of $SU(2)_R$ and the $(n+1)^\mathrm{th}$ antisymmetric representation of $SU(2n-1)$. Once again, when combined with the anti-instanton, baryons and anti-baryons, these exactly form the states we observed from the unbalanced nodes in the 3d quiver.

\section{$SU(n)_{\pm1/2}$ with $N_f=2n-1$ flavors} \label{sec:2nm1}
This is the third $E_4$ sequence.

For this case, the corresponding subgroup of $E_4$ is $SU(3)\times SU(2)\times U(1)$ which generalizes to $SU(2n-1)\times SU(2)\times U(1)$. This implies that there are 2 unbalanced nodes which can be determined by looking at the difference between the affine $E_4$ Dynkin diagram and the $SU(3)\times SU(2)$ Dynkin diagram. This fixes the 3d quiver uniquely.
The Higgs branch at infinite coupling is given by the Coulomb branch of the $3d$ quiver:\\
\bea \label{eq:2nm1}
\CH_\infty \left( \node{}{SU(n)_{\frac{3}{2}}}-\sqnode{}{2n-1}  \right)  = \mathcal{C}^{3d} \left ( \begin{tikzpicture}[baseline={([yshift=-.5ex]current bounding box.center)}]
\node[anchor=south west] at (0,0)  {$\node{}{1}- \node{}{2}-\cdots- \node{}{n-2}-\node{\bver{1}{}}{n-1}-\node{\bver{}{1}}{n-1}- \node{}{n-2} -\node{}{2}-\node{}{1}$} ;  
\node[anchor=south west] at (4.3,1.15) {$\node{}{1}$}; 
  \draw[black] (4.1,1.3)--(4.4,1.65); 
    \draw[black] (4.95,1.3)--(4.7,1.65); 
\end{tikzpicture} \right )~.  \eea
The quaternionic Higgs branch dimension at infinite coupling of the quiver on the left hand side can be computed from the Coulomb branch dimension of the quiver on the right hand side:
\be \label{dimSU2nm1SU2U1}
\dim \CH_\infty = n^2-n+2
\ee
This increases the finite coupling Higgs branch that has dimension $n^2-n+1$ by adding $1$ new flat direction.  There is a classical global symmetry $U(2n-1) \times U(1)_I \cong SU(2n-1) \times U(1)_B \times U(1)_I $ with rank $r=2n$.

The global symmetry at infinite coupling can be read off from the above quiver, after removing the blue nodes: this operation results with a flavor symmetry $SU(2n-1) \times SU(2) \times  U(1)$, which has rank $2n$, as expected from finite coupling.

The HWG is conjectured to be
\be \label{SU2nm1SU2U1}
\PE \left[ \sum_{i =1}^{n-1} \mu_i \mu_{2n-i-1} t^{2i} + (\nu^2 + 1) t^2 + \nu (\mu_{n-1}q + \mu_n q^{-1})t^n - \nu^2 \mu_{n-1} \mu_n t^{2n} \right]~.
\ee
Indeed, for $n=2$, we recover the Hilbert series of the minimal nilpotent orbit of $E_4 = SU(5)$, written in terms of the highest weight of $SU(3)\times SU(2)\times U(1)$.  For $n=3$, we compute the Coulomb branch Hilbert series of the quiver of the right hand side of \eref{eq:2nm1} up to order $t^6$ using the Coulomb branch formula \cite{Cremonesi:2013lqa}, and find the agreement with the Hilbert series obtained from the HWG \eref{SU2nm1SU2U1}.  
Explicitly, the unrefined Hilbert series for $n=3$ is
\be
\begin{split}
& \sum_{u=-\infty}^\infty\,\, \sum_{ v=-\infty}^\infty \,\, \sum_{w_1 = -\infty}^\infty\,\, \sum_{w_2 = -\infty}^\infty\,\, \sum_{w_3 = -\infty}^\infty \,\, \sum_{b_1 \geq b_2 > -\infty} \,\, \sum_{a_1 \geq a_2 > -\infty} {\blue \delta(v)} \\
& \qquad t^{\sum_{i=1}^2|u-a_i|+\sum_{i=1}^2|v-b_i|}  t^{\sum_{i=1}^2|w_1-a_i|+\sum_{j=1}^2|w_2-b_j|+|w_3-w_1|+|w_3-w_2|} t^{\sum_{i,j=1}^2 |a_i-b_j|}  \\
& \qquad \times t^{-2|a_1-a_2|} t^{-2|b_1-b_2|} P_{U(2)}(t; a_1,a_2) P_{U(2)}(t; b_1,b_2)  \\
& \qquad \times P_{U(1)}(t; u) P_{U(1)}(t; v) {\blue (1-t^2)} \prod_{i=1}^3 P_{U(1)}(t; w_i)~.
\end{split}
\ee
where the factors in the second line denote the contributions from the bi-fundamental hypermultiplets; those in the third line denote the contributions from the $U(2)$ gauge groups whose fluxes are denoted by $\vec a$ and $\vec b$; and those in the fourth line denote the contributions from the $U(1)$ gauge groups whose fluxes are denoted by $u, v, w_1, w_2, w_3$ (where $w_1, w_2, w_3$ correspond to those in the loop and $u$, $v$ correspond to those at the left and right ends of the quiver).  The function $P_{U(m)}(t; n_1, \ldots, n_m)$ is given by (A.2) of \cite{Cremonesi:2013lqa}. The removal of an overall $U(1)$ is denoted in blue (namely, multiplying by the factor $(1-t^2)$ and setting $v=0$); in the above, this is done from one of the $U(1)$ nodes at the left or the right end of the 3d quiver. Evaluating the summations, we find that
\be
1 + 28 t^2 + 40 t^3 + 380 t^4 + 820 t^5 + 3656 t^6+ \ldots~.
\ee
Alternatively, the removal of an overall $U(1)$ can be done from the other node; for example, if this is done from one of the nodes labelled by $n-1$, we simply change the summation to be
\be
\sum_{u=-\infty}^\infty\,\, \sum_{v=-\infty}^\infty \,\, \sum_{w_1=-\infty}^\infty \,\, \sum_{w_2 = -\infty}^\infty\,\, \sum_{w_3 = -\infty}^\infty \,\, \sum_{b_1 \geq {\blue b_2 =0}} \,\, \sum_{a_1 \geq a_2 > -\infty}
\ee
with {\it no} delta function for $v$. This yields the same Hilbert series as above.

Another non-trivial test of \eref{SU2nm1SU2U1} is to derive the dimension of $\CH_\infty$ from this function, in the way described in \cite[sec. 4.3]{Hanany:2015hxa}, and compare it with \eref{dimSU2nm1SU2U1}. To derive the former, we use the following data: 
\bi
\item The HWG dimension of \eref{SU2nm1SU2U1} is $(n-1)+2+2-1 = n+2$.  
\item The irrep structure of $SU(2n-1) \times SU(2)$ that appears in the above HWG is $[m,m, \ldots, m]_{SU(2n-1)}[m]_{SU(2)}$, with $n\neq 0$.  The dimension of such a representation is a polynomial in $m$ of degree ${2n-1 \choose 2}+1 = 2n^2-3n+2$.
\ei
The sum of the above two quantities is $(n+2)+(2n^2-3n+2) = 2n^2-2n+4$.  This is the expected complex dimension of $\CH_\infty$ as derived from \eref{SU2nm1SU2U1}. Indeed, it is in agreement with the quaternionic dimension given by \eref{dimSU2nm1SU2U1}.  The conjectured HWG thus passes this test.

\subsection{5d analysis}

The 5d analysis follows similarly to the previous cases. There are two contributions at the 1-instanton order. The first is in a singlet of $SU(2n-1)$ and provides the conserved current that, with the anti-instanton, forms the enhanced $SU(2)$. The second one is in the $\bold{n+1}$ of $SU(2)_R$ and the $n^\mathrm{th}$ antisymmetric representation of $SU(2n-1)$. Once again, when combined with the anti-instanton, baryons and anti-baryons, these exactly form the states we observed from the unbalanced nodes in the 3d quiver.

\section{Connecting the 5d gauge theory and 3d quiver} \label{sec:5dweb}

Throughout this article we have used the Coulomb branches of 3d quivers to realize the Higgs branches of 5d gauge theories. It may at first seem mysterious why such a technique should work. Yet, with help from the magic of $8$ supercharges and 3d mirror symmetry, it can be physically motivated as follows. We can consider reducing the 5d SCFT, which is the UV completion of the gauge theory, to 3d on a torus. Because of the amount of supersymmetry, the Higgs branch does not receive quantum corrections, and thus will be the same one as in 5d. Many 3d theories have mirror duals where the Coulomb branch of the mirror dual realize the originals Higgs branch.

Of course generically there is no guarantee that the mirror dual will be Lagrangian. However, in this case, we have good reasons to suspect that this should hold at least for some cases. This follows due to the results of \cite{Benini:2010uu,Benini:2009aa}. Particularly, it was argued in \cite{Benini:2009aa} that a specific class of 5d SCFTs, those described by an intersection of D5-branes, NS5-branes and $(1,1)$ 5-branes, reduces to A type class S theories when compactified on a circle. In \cite{Benini:2010uu}, it was argued that reducing A type class S theories on a circle to 3d leads to 3d SCFTs possessing Lagrangian mirrors whose shape is a three legged quiver of unitary groups\footnote{In each of these quivers an overall $U(1)$ needs to be modded out.}. Furthermore, 5d gauge theories with sufficiently many flavors have 5d SCFTs belonging to this class. Therefore, these results strongly suggest that such a method should work at least for gauge theories with enough matter. For $SU(n)$ groups with fundamental matter, this turns out to be $N_f > 2n$. 

\begin{figure}
\center
\includegraphics[width=0.95\textwidth]{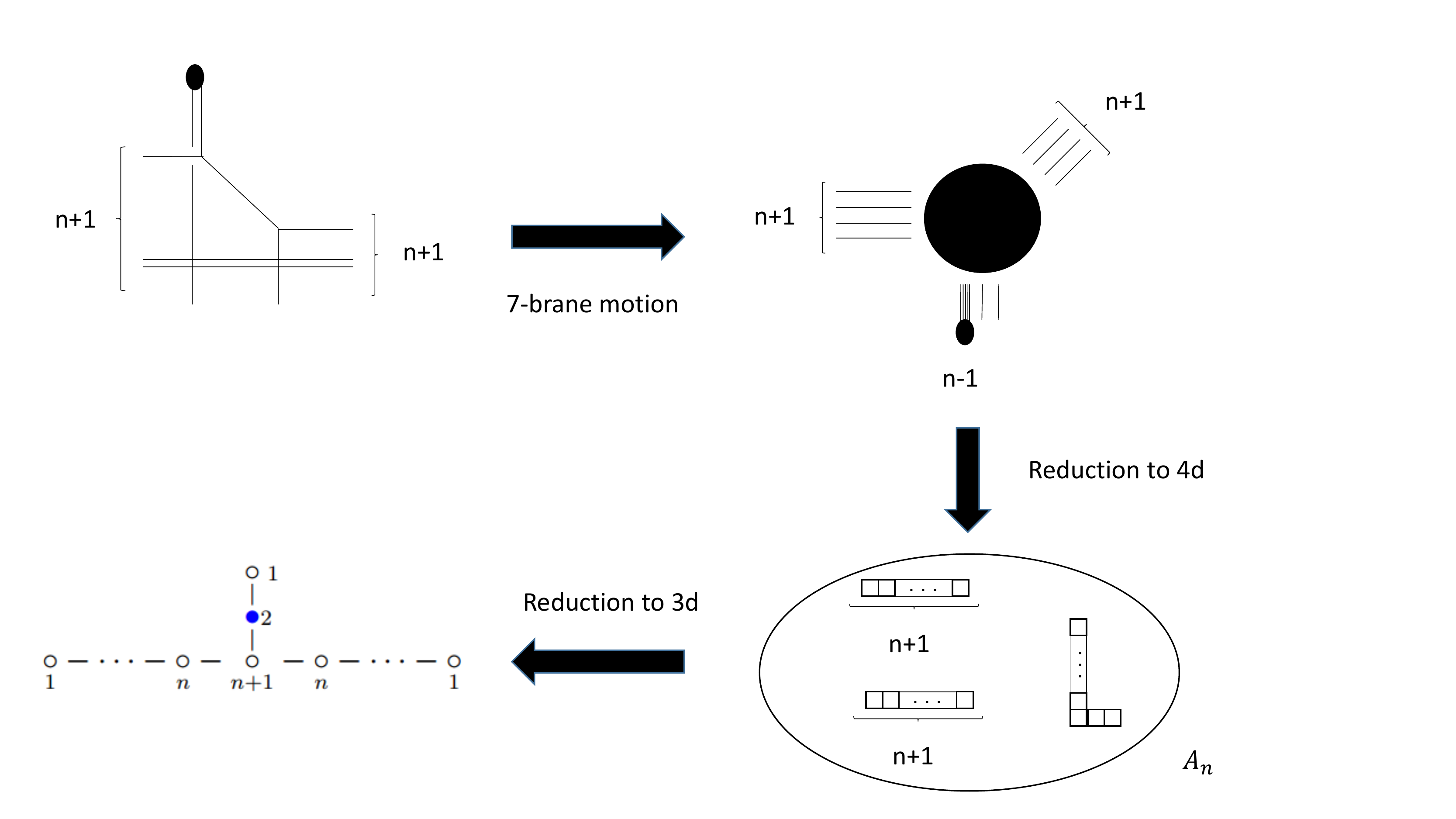} 
\caption{Deriving the 3d quiver using the brane web. At the top left is the 5d brane web describing an $SU(n)_{\frac{1}{2}}+(2n+1)F$ gauge theory. Moving the top 7-brane to the bottom leads to the web on the top right. This one is in the form of \cite{Benini:2009aa} and so reduces to the class S theory shown on the bottom right when compactified on a circle to 4d. Further compactification on a circle to 3d leads to the 3d quiver.}
\label{WebtoQuiver}
\end{figure}

This method can then also be used to derive the 3d quivers in those cases. An example of this is shown in figure \ref{WebtoQuiver} for the 5d gauge theory $SU(n)_{\pm \frac{1}{2}}+(2n+1)F$ which is related to the 4d $R_{(0,n+1)}$ theory of \cite{Chacaltana:2010ks}. The other cases can also be derived similarly. The cases with $2n-1 \leq N_f \leq 2n$ require a slightly different approach.

\begin{figure}
\center
\includegraphics[width=0.95\textwidth]{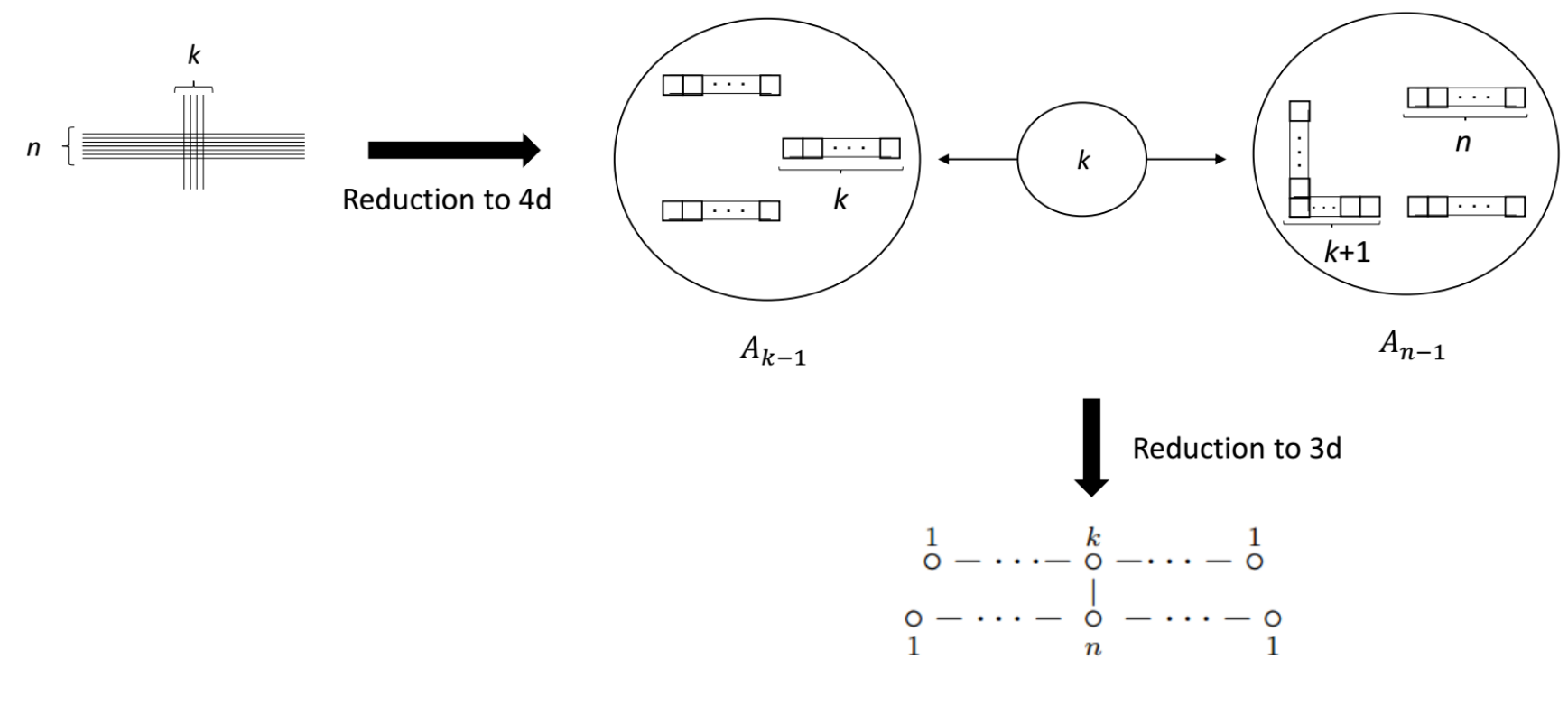} 
\caption{Starting with the 5d SCFT described by the brane web in the top left, and reducing to 4d, it is conjectured by \cite{Ohmori:2015pia} that we get the 4d theory on the top right, where here we take $n>k$. The small circle denotes an $SU(k)$ group and the arrows indicates it is gauging an $SU(k)$ global symmetry in both class S theories. Reducing further to 3d leads to the mirror quiver on the bottom.}
\label{WebtoQuiver1}
\end{figure}

 For the $N_f = 2n$ case can use the results of \cite{Ohmori:2015pia} together with some properties of 3d quivers to derive the quivers in an alternative method. Specifically, \cite{Ohmori:2015pia} studied the 4d reduction of the brane webs of the type shown in the top left of figure \ref{WebtoQuiver1}.  They conjectured that this reduces to the IR free 4d theory shown in the top right of figure \ref{WebtoQuiver1} (here we assume that $n>k$). Now consider the 3d reduction of this theory. Each class S theory can be reduced to a mirror star shaped quiver which are now connected via gauging part of the global symmetry on the Coulomb branch. The symmetry in question is an $SU(k)$ group generated by the tail $U(1) \times U(2) \times ... \times U(k-1)+kF$. It is known that connecting two quivers in this way act as a delta function identifying the two $SU(k)$ global symmetries associated with the flavors. This plays an important role in the 3d mirror quiver and class S correspondence. In our case this implies that the theory we get in 3d is build from two star shaped quivers with one leg removed that have been adjoined along the removed leg. This gives the quiver shown in the bottom of figure \ref{WebtoQuiver1}. For the case of $k=2$ this reduces to the quiver for $SU(n)_0+2nF$. The other cases can also be cast in this form via 7-brane motion where some of the 5-branes are forced to end on the same 7-brane. This corresponds in 4d to changing the maximal puncture to a smaller type. This just changes the leg associated with it to the one associated with the puncture as is ordinary in \cite{Benini:2010uu}. 

\begin{figure}
\center
\includegraphics[width=0.95\textwidth]{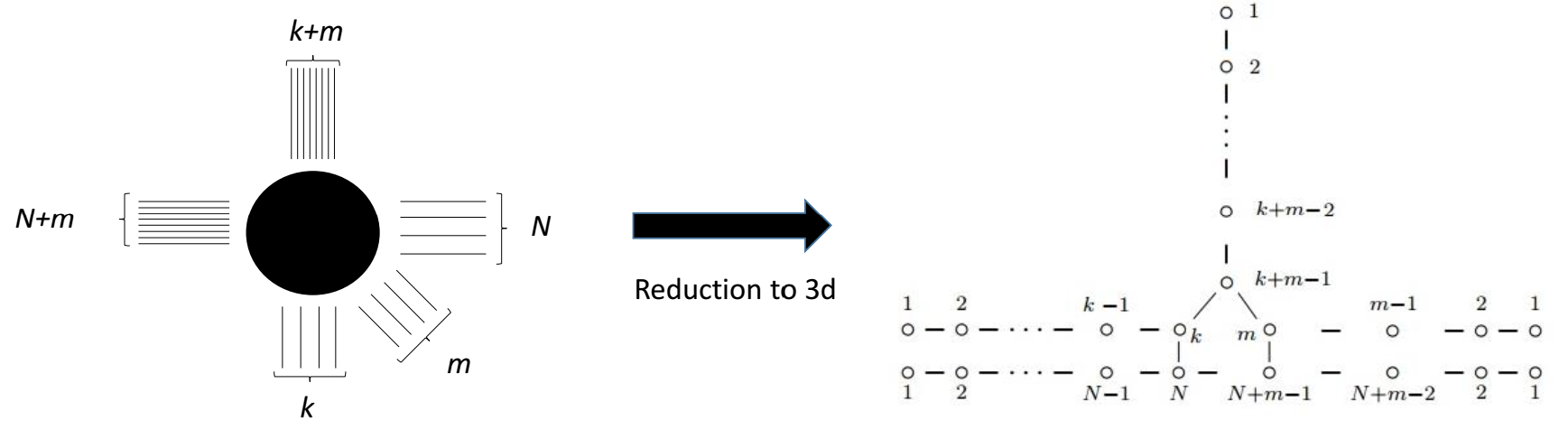} 
\caption{We conjecture that reducing the 5d SCFT described by the brane system on the left leads to a 3d theory with a mirror dual given by the quiver on the right.}
\label{WebtoQuiver2}
\end{figure}

The cases associated with $N_f = 2n - 1$ cannot be tackled using known results, at least to our knowledge. However, having found promising 3d quivers whose Coulomb branch seems to describe the gauge theory Higgs branch, it is tempting to use this logic in reverse and conjuncture that the 5d SCFTs reduce to 3d theories with these mirror duals. A non-trivial additional test of this is that the dimension of the Higgs branch of the 3d quiver agrees with the Coulomb branch dimension of the 5d SCFT. We can refine the statement to one, which claims that the 5d SCFT described by the web on the left of figure \ref{WebtoQuiver2}, when reduced to 3d, has a mirror dual given by the quiver on the right of figure \ref{WebtoQuiver2}. We can again perform the same checks on this proposal, particularly matching the dimensions of the Higgs and Coulomb branches. 

When $m=k=1, N=n-1$, this reduces to the quiver for $SU(n)_{\frac{1}{2}}+(2n-1)F$. The other cases can also be cast in this form via 7-brane motion where some of the 5-branes are forced to end on the same 7-brane. Again this corresponds in 3d to changing the leg associated with the maximal puncture to a smaller type, the only difference is in the legs associated with the $k+m$ and $N+m$ collection of 5-branes whose associated puncture is the ordinary one for the collection once the 7-brane with the maximum number of 5-branes ending on it is removed.

\acknowledgments
We would like to thank Stefano Cremonesi for the collaboration at the early stages of this work.  The research of N.~M.~ is supported by the INFN.  A.~H.~ and G.~Z.~ thank the Workshop on ``SuperConformal Field Theories in Four or More Dimension" and the Aspen Centre for Physics for their kind hospitality, where this project was initiated. A.~H.~ and N.~M.~ gratefully acknowledge the Pollica Summer Workshop 2017 (partly supported by the ERC STG grant 306260) and, with G.F., the Simons Summer Workshop 2017, where significant progress of this project has been made. A.~H.~ would like to thank the National Taiwan University and the National Center for Theoretical Science in Taiwan for their kind hospitality during the final stages of this work. A.~H.~ is supported in part by an STFC Consolidated Grant ST/J0003533/1, and an EPSRC Programme Grant EP/K034456/1. G.~Z.~ is supported in part by World Premier International Research Center Initiative (WPI), MEXT, Japan.

\bibliographystyle{ytphys}
\bibliography{ref}

\end{document}